\documentclass[prl,twocolumn,nofootinbib,superscriptaddress]{revtex4-2}

\usepackage{preamble}

\usepackage{xcolor}
\usepackage{hyperref}
\definecolor{purple}{HTML}{4B2E83}
\hypersetup{colorlinks=true,allcolors=blue}

\begin{document}

\title{Tracing source imprints on the gravitational direct wave
}

\author{Andrew Laeuger\,\orcidlink{0000-0002-8212-6496}}
\email[Contact author: ]{alaeuger@caltech.edu}
\author{Colin Weller\,\orcidlink{0000-0001-5173-5638}}
\affiliation{Burke Institute for Theoretical Physics and Theoretical Astrophysics 350-17, California Institute of Technology, Pasadena, CA 91125, USA}

\author{Ling Sun\,\orcidlink{0000-0001-7959-892X}}
\affiliation{OzGrav-ANU, Centre for Gravitational Astrophysics, Research School of Physics and Research School of Astronomy \& Astrophysics, The Australian National University, ACT 2601, Australia}

\author{Aleyna Koro\,\orcidlink{0009-0009-5530-4574}}
\affiliation{School of Physics and Center for Relativistic Astrophysics, Georgia Institute of Technology, Atlanta, GA 30332, USA
}

\author{Patrick P. Chen}
\affiliation{Burke Institute for Theoretical Physics and Theoretical Astrophysics 350-17, California Institute of Technology, Pasadena, CA 91125, USA}

\author{Sizheng Ma\,\orcidlink{0000-0002-4645-453X}}
\affiliation{Perimeter Institute for Theoretical Physics, Waterloo, ON N2L2Y5, Canada}

\author{Yanbei Chen\,\orcidlink{0000-0002-9730-9463}}
\affiliation{Burke Institute for Theoretical Physics and Theoretical Astrophysics 350-17, California Institute of Technology, Pasadena, CA 91125, USA}

\date{\today}

\begin{abstract}
    The direct wave component of black hole (BH) ringdown, revealed through quasinormal-mode filters, carries a non-quasinormal-mode imprint of the plunging perturber's dynamics. However, the connection between the source trajectory and the observed direct wave remains incompletely understood. We first show that the full waveform filtered by the frequency-dependent black hole response gives a substantially more localized kernel than the standard rational filter approach. We then show that analytically continuing the Teukolsky source variables to complex coordinates allows for a semianalytic direct wave which closely reproduces numerical integration throughout the entire plunge. Our generalized steepest-descent method clarifies how coherent emission shapes the imprint of source dynamics on the observed direct wave beyond the light-cone picture.
\end{abstract}

\maketitle

\ssec{Introduction} 
The gravitational ringdown following a compact-binary coalescence is well understood at late times, when the signal is dominated by quasinormal modes (QNMs) \cite{QNMs1, QNMs2, QNMs3, QNMs4, QNMs5} and power-law tails \cite{Tails1, Tails2, Tails3}. Recent work has increasingly focused on the early ringdown, including quadratic QNMs \cite{QQNM1, QQNM2, QQNM3, QQNM4, QQNM5, QQNM6, QQNM7, QQNM8}, dynamical excitation of linear QNMs \cite{Marina1, Marina2, KerrDynamical}, and the prompt response of the black-hole Green's function \cite{Prompt1, Prompt2, Prompt3, Prompt4}.

Upon application of frequency-domain filters which remove the QNMs from the early ringdown \cite{MaFilters}, a ``direct wave'' is revealed, which is canonically associated with radiation propagating from the infalling particle to a distant observer as the particle plunges toward the horizon \cite{SizhengPaper1}. This component offers a probe of how the plunging source is imprinted on the observed signal, though whether it conveys properties of the remnant Kerr horizon remains an active question \cite{SizhengPaper2, AdrianPaper1, AdrianPaper2, KankaniPaper1, KankaniPaper2, DirectWaveEvolution, Dyer:2026sox, Shi2026}. In particular, the Matsubara (or ``horizon'') frequencies $\omega_\text{MB}^{(n)}$ associated with the direct wave are screened at future null infinity by the Teukolsky source or the black-hole transmissivity $B_\text{in}(\omega)^{-1}$ \cite{KerrDynamical, Kuntz:2026xep, SizhengPaper1, Kubota:2026qjh, Nakamoto:2026rah, SizhengPaper3}.
Some recent interpretations in the literature which characterize the direct wave by a single horizon frequency are incomplete: the direct wave is characterized by a time-dependent frequency that evolves rapidly before approaching a universal frequency determined exclusively by the horizon properties \cite{DirectWaveEvolution, Dyer:2026sox}, in close connection with dynamical QNM excitation \cite{KerrDynamical, SizhengPaper3, Kubota:2026qjh, Nakamoto:2026rah, Marina1, Marina2}. However, the exact map between the source and the evolution of the direct wave throughout the entire plunge, going beyond just the near-horizon region, is not yet completely understood.

In this letter, we seek to understand how each part of the particle trajectory contributes to the outgoing waveform, and which choice of linear filter applied to the outgoing wave can best allow us to extract information at different stages of the trajectory. We identify a previously unappreciated filter, which we dub the ``response filter'', which succeeds in localizing each moment of the trajectory in the corresponding waveform, as we demonstrate via the structure of the filtered waveform kernels.

Furthermore, we show that a generalized method of steepest descent is a viable approach for computing the direct wave under this choice of filter. Prior steepest-descent calculations, which use near-horizon approximations and simplify the frequency dependence entering the saddle-point conditions \cite{SizhengPaper1}, do not perform well near the light-ring (LR) crossing, where source contributions evolve rapidly \cite{Kubota:2026qjh, Shi2026}. We instead analytically continue the plunge trajectory and homogeneous Teukolsky solutions to complex coordinates, retaining the full frequency dependence of each source term to obtain a semianalytic waveform that closely matches direct numerical integration throughout the entire plunge.

The generalized steepest-descent method also provides a framework for moving beyond the limited light-cone intuition in connecting the complete source trajectory to the observed waveform. 
We show that the complex saddle time describes the properties of coherent radiation emitted along the trajectory, and that while localized modifications to the source register near their light-cone arrival, temporally extended modifications appear in the waveform with a lag given by this coherent emission time. Our results refine the picture of the direct wave beyond its current understanding as a sequence of wave packets propagating exclusively along the light-cone.

In this work, we use geometrized units $G=c=M=1$, where $M$ is the central black hole mass. We also focus on plunges from the prograde innermost stable circular orbit (ISCO) for black-hole spins $\chi=\{0.5,0.7,0.9\}$, though the results and methods developed here can be extended to more generic trajectories. The energy and angular momentum are set to their ISCO values, and the initial radial coordinate is set just inside $r_\text{ISCO}$. The trajectory then proceeds along a geodesic, with equations of motion appearing in the Supplemental Material.

\ssec{Direct wave extraction by linear filters} The direct wave is typically extracted from plunge and binary coalescence waveforms $Z(u)$ via the application of a linear filter $\tilde{\mathcal{F}}(\omega)$. The aim of the QNM filtering program is to identify filters which remove the QNMs such that $\tilde{z}(\omega)=\tilde{\mathcal{F}}(\omega)\tilde{Z}(\omega)$ reveals the underlying structure of the source. The most common choice is the rational QNM filter \cite{MaFilters, SizhengPaper2}, defined for a set of modes $\{\omega_\text{QNM}\}$ by 
\begin{equation}
    \tilde{\mathcal{F}}_\text{rat}(\omega)=\prod_{\{\omega_\text{QNM}\}}\frac{\omega-\omega_\text{QNM}}{\omega-\omega_\text{QNM}^*}\frac{\omega+\omega_\text{QNM}^*}{\omega+\omega_\text{QNM}}.
\end{equation}
Previous studies have generally assumed that in the analytic particle plunge model, the rationally-filtered Teukolsky source integrand admits a decomposition into a slowly-varying prefactor and rapidly varying phase, motivating the method of steepest descent to compute the integral. By retaining only the leading contributions to the source in the near-horizon region and placing all of the $\omega-$dependence of those coefficients and filters into the slowly-varying prefactor, the approximate saddle frequency and observer time are given by:
\begin{equation}
    \omega_G(t)=m\frac{\beta\Omega_H+\Omega}{1+\beta}-i\frac{2\beta\kappa_H}{1+\beta},\quad u(t)=t-r_*,
    \label{eq: omega_G}
\end{equation}
where $\beta=-dr_*/dt$, $\Omega=d\phi/dt$, $\Omega_H=\chi/(2r_+)$ is the horizon angular frequency, $r_+=1+\sqrt{1-\chi^2}$ is the outer horizon radius, and $\kappa_H=\sqrt{1-\chi^2}/(2r_+)$ is the surface gravity, with $\chi$ the dimensionless black-hole spin. At late times, $\omega_G$ approaches the first Matsubara frequency, $\omega_\text{MB}^{(1)}=m\Omega_H-i\kappa_H$.

This simplified model yields a complex frequency which evolves throughout the plunge according to the trajectory dynamics. However, it is not without problems -- recent calculations have pointed out that the isolated prefactor may violate the assumption of slow variance \cite{Kubota:2026qjh, Shi2026}, and the near-horizon approximations used in Eq. \eqref{eq: omega_G} do not hold well for the majority of the plunge. 
Furthermore, the rational filters are limited in their ability to map individual contributions to the direct wave -- Wang and Ma have shown that the anti-causal poles extract waves emitted from within the light-ring to substantially early times, distorting the wave \cite{SizhengPaper3}. 

To address these problems with the  existing approach, we propose a new choice of QNM filter, which admits an improved steepest-descent treatment. In Ref. \cite{KerrDynamical}, the outgoing wave was decomposed as $\tilde{Z}_{lm}(\omega) =\tilde{\mathcal{G}}_{\text {BH},lm}(\omega) \tilde{\mathcal{I}}_{lm}(\omega)$, so that in the time domain, it forms the convolution 
\begin{equation}
    Z_{lm}(u) = \int \mathcal{G}_{\text{BH},lm}(u-u') \mathcal{I}_{lm}(u') du'.
    \label{eq: Z_lmt}
\end{equation}
Here, $\tilde{\mathcal{G}}_{\text{BH},lm}$ is the causal response function of the black hole\footnote{We often suppress $(l,m)$ indices, taking $l=m=2$ throughout.} and $\mathcal{I}_{lm}$ is a Teukolsky-source-dependent integral along the plunge trajectory of a particle, of the form \begin{equation}
    \mathcal{I}_{lm}(u) = \int \frac{d\omega}{2\pi} e^{-i\omega u}\int dt~ \mathcal{S}_{lm\omega}(t),
     \label{eq: I_lm}
\end{equation}
where $t$ is the coordinate time along the trajectory.
\begin{figure}[t]
    \centering
    \includegraphics[width=0.99\linewidth]{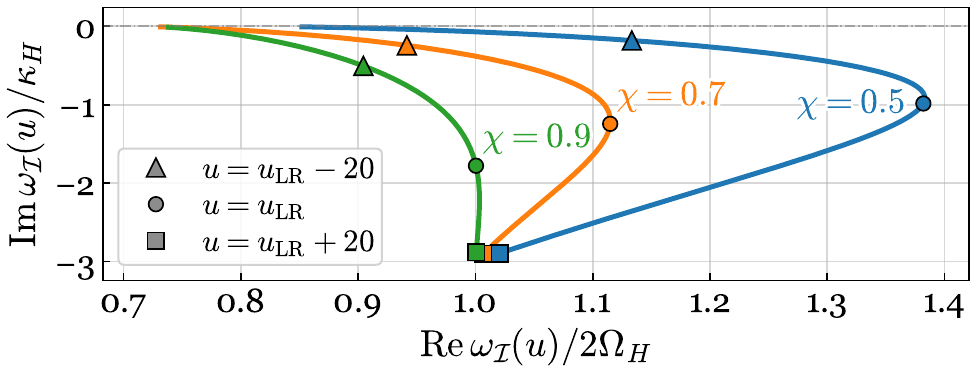}
    \caption{The instantaneous frequency of the source $\mathcal{I}_{22}(u)$. The frequency evolves rapidly near $u_\text{LR}=t_\text{LR}-r_{*,\text{LR}}$ before approaching $\omega_\text{MB}^{(3)}$ at late times, regardless of spin.}
    \label{fig:source freq}
\end{figure}
At late times, the instantaneous frequency of $\mathcal{I}_{22}(u)$ approaches $\omega_\text{MB}^{(3)}$ -- an imprint of horizon properties noted in \cite{KerrDynamical} and shown explicitly in Figure \ref{fig:source freq}.

The form of Eq. \eqref{eq: Z_lmt} motivates a ``response filter'' of the simple form $\tilde{\mathcal{F}}_\text{resp}\equiv\tilde{\mathcal{G}}^{-1}_\text{BH}$: since $\tilde{\mathcal{G}}_\text{BH}$ carries poles at the QNMs and zeroes at the $n\geq 3$ Matsubara modes in the complex frequency domain \cite{KerrDynamical, Kubota:2026qjh, Nakamoto:2026rah, Kuntz:2026xep}, $\tilde{\mathcal{F}}_\text{resp}$ naturally filters out these features. The resulting direct wave is just $z_\text{resp}(u)=\mathcal{I}(u)$: its complex frequency evolution is exactly that of the source itself, rather than being distorted through its convolution with the BH response function and rational filters.

\begin{figure*}[t]
    \centering
    \includegraphics[width=0.95\linewidth]{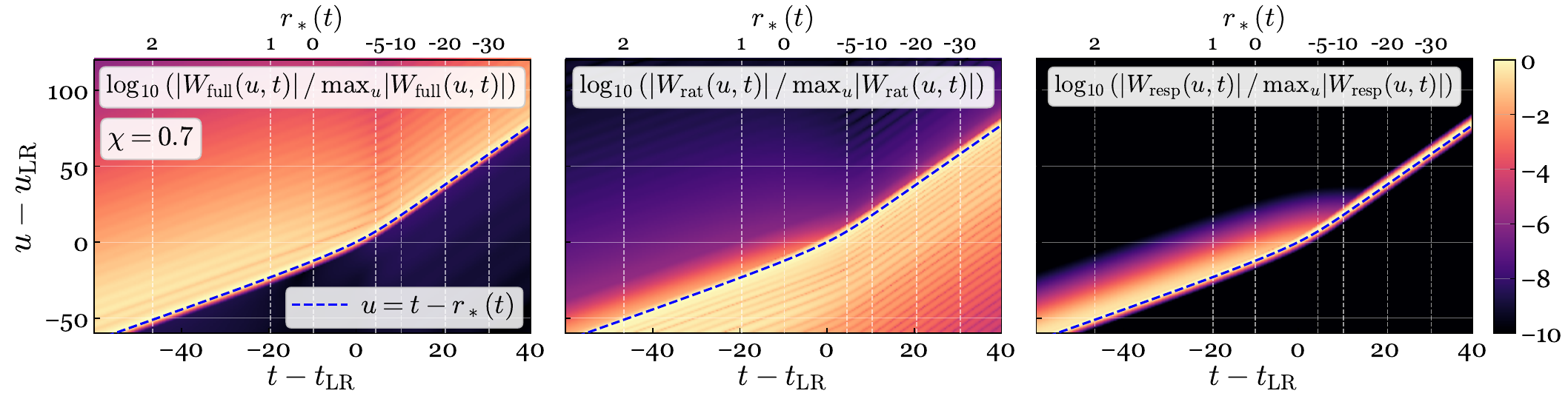}
    \caption{The magnitude kernel $W(u,t)$, normalized by its maximum over $u$ for each value of $t$. The unfiltered (``full") kernel displays a wake of beating QNMs which trails behind the peak response on the light-cone, while the rationally filtered kernel flips the QNM beating into a leading wake. In both cases, for a given $t$, the kernel is spread out over a wide range of $u$; when the response filter is applied instead, the kernel becomes substantially more localized, with the peak remaining on the light-cone but the trailing wake decaying sharply at a rate of $3\kappa_H$.}
    \label{fig:filtered kernels}
\end{figure*}

\ssec{Structure of the filtered kernels} We now evaluate whether the response filter improves upon the rational filters in its mapping between the source dynamics and resulting direct wave.
Suppose we add a profile function $\xi(t)$ along the trajectory, namely $\mathcal{S}_{lm\omega}(t) \rightarrow  \mathcal{S}_{lm\omega}(t) \xi(t) $, constructed to isolate a particular segment of the plunge. A useful tool to understand the imprint of this segment on the direct wave is the function
\begin{equation}
\frac{\delta z (u)}{\delta \xi (t)} \!=\!\int \frac{d\omega}{2\pi}e^{-i\omega u} \tilde{\mathcal{F}}(\omega)  \tilde{\mathcal{G}}_{\rm BH}(\omega) \mathcal{S}_{lm\omega}(t) \equiv W(u,t). 
\label{eq: kernel}
\end{equation}
In this language, $W(u,t)$ is the transfer function, or kernel, from a modification of the source at $t$ of the trajectory to the filtered wave at $u$.  Without any filter, the kernel at fixed $t$ is substantially broadened by the QNM ringing of the central black hole.

In Figure \ref{fig:filtered kernels}, we compute Eq. \eqref{eq: kernel} for the case of no filter ($\tilde{\mathcal{F}}=1$), the response filter, and the rational filter where the set $\{\omega_\text{QNM}\}$ contains the $l=2$, $m=\pm2$, $n\leq2$ modes with $\mathrm{Re}(\omega_\text{QNM})>0$, a set motivated by the finding of \cite{KerrDynamical, Nakamoto:2026rah} that higher overtones are closely related to the late-time behavior of $\mathcal{I}(u)$. Each slice of the kernel along a fixed $t$ is normalized by its maximum over $u$. 

The unfiltered kernel $W_\text{full}$ features a peak on the light-cone, followed by a trailing wake whose magnitude oscillates slightly on top of a decaying trend due to beating between QNMs. When the rational filters are applied, the wake now leads the peak on the light-cone due to the introduction of acausal poles, as discussed in \cite{SizhengPaper3}. However, when the response filter is applied, the kernel becomes substantially more localized in time, with its support on the real $t$ axis concentrated heavily on the light-cone. The trailing wake now decays at $3\kappa_H$, set by the pole in $R_\text{in}$ at the $\omega_\text{MB}^{(3)}$ (see the Supplemental Material), far exceeding that of the fundamental QNM. This behavior suggests that the response filter may address at least one of the rational filter shortcomings, generating a more straightforward map between any given segment of the direct wave and the corresponding piece of the plunge which sourced it.

\begin{figure}
    \centering
    \includegraphics[width=\linewidth]{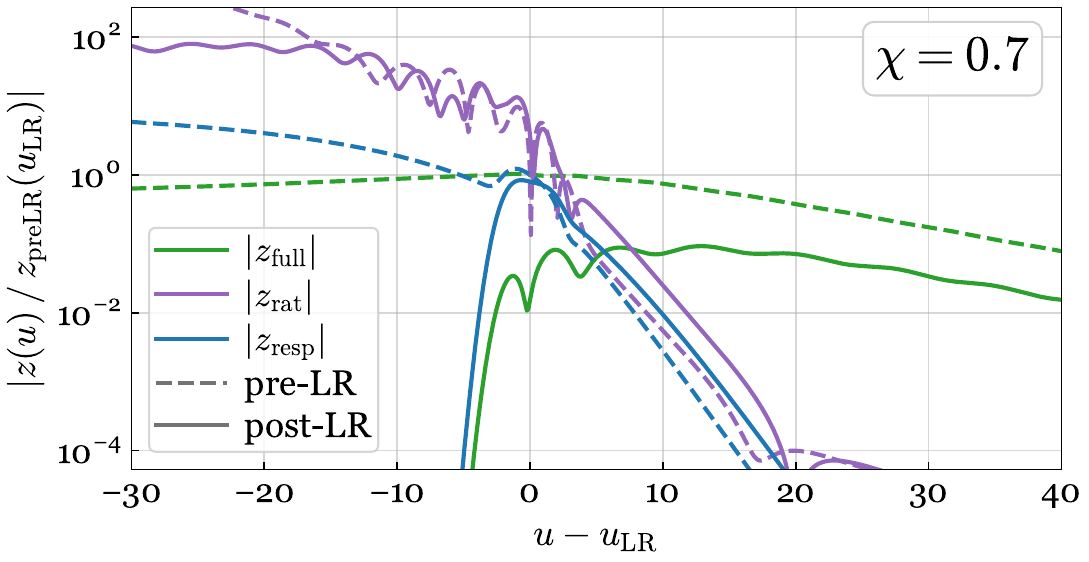}
    \caption{Waveforms contributed from the segments of the trajectory before and after the light-ring crossing, either unfiltered or filtered with the rational or response filter. The post-LR contribution is hidden by the pre-LR contribution in the unfiltered, but the rational filter and response filter both boost the relative magnitude of the post-LR portion.}
    \label{fig: lr split waveforms}
\end{figure}

The imprint of this greater localization becomes more apparent when we integrate the kernel over $t$ with a profile $\xi(t)$ imposed. We choose $\xi(t)$ as a Planck-taper window which smoothly turns on or off at the light-ring crossing time $t_\text{LR}$, isolating the pre-LR and post-LR contributions to the observed waveform in Figure \ref{fig: lr split waveforms}.

When no filter is applied, the portion of the waveform sourced inside the light-ring is buried beneath the portion sourced externally -- after the post-LR waveform ramps up, it still lies a factor of $\sim 5$ below the pre-LR windowed waveform, in agreement with the findings of \cite{Cheung:2026gfd}. Once the rational filters are applied, the post-LR waveform becomes comparable to or greater than its pre-LR counterpart when $u>u_\text{LR}\equiv t(r_\text{LR})-r_{*,\text{LR}}$, though the post-LR waveform still has support for $u<u_\text{LR}$ due to the introduced anti-causal poles and displays beating with those reversed QNMs in the vicinity of $u_\text{LR}$.

With the response filter, however, the separation between the pre- and post-LR contributions is clearly defined: causality is evident, with the post-LR waveform turning on\footnote{The sharp ramp-up prior to $u_\text{LR}$ is dictated by a Gaussian filter used to regularize the $\omega$ integral in Eq. \eqref{eq: kernel}.} and becoming the dominant piece of the full plunge waveform right at $u_\text{LR}$. Thus, the response filter readily isolates the source dynamics over each stretch of the plunge trajectory, though it is important to note that the signal-to-noise ratio (SNR) accumulated over the late plunge is relatively small due to the decay of the source. 
 
\ssec{Steepest-descent with analytic continuation} 
We now demonstrate a generalized steepest-descent approximation to the response-filtered waveform. The BH response and Teukolsky source functions are \cite{MinoReview, SasakiReview}
\begin{subequations}
\begin{align}
    &\tilde{\mathcal{G}}_{\text{BH},lm} \equiv (2i\omega B_\text{in}(\omega))^{-1},
    \\
    &\mathcal{S}_{lm\omega} \equiv e^{i\omega t-im\phi(t)}
    \Bigl[
        (A_{nn0}+A_{\bar m n0}+A_{\bar m\bar m0})R_\text{in}
        \nonumber\\
    &\qquad
        -(A_{\bar m n1}+A_{\bar m\bar m1})R_\text{in}'
        + A_{\bar m\bar m2}R_\text{in}''
    \Bigr].
    \label{eq: S_lmw}
\end{align}
\end{subequations}
Here, $R_\text{in}$ is the homogeneous Teukolsky solution which is purely ingoing at the horizon with ingoing amplitude of $B_\text{in}$ at infinity, primes denote $\partial/\partial r$, and the source coefficients $A_j$ are given in the Supplemental Material -- the source dependence on the trajectory $(r(t),\phi(t))$ and $\omega$ is notationally suppressed. The transformation from $Z_{lm}$ to the Newman-Penrose scalar $\Psi_4$ is given in \cite{MinoReview,SasakiReview}.

To resolve the problems identified for the steepest-descent calculation which produced Eq. \eqref{eq: omega_G}, we apply the method of steepest descent across the entire plunge in its most generic form. This requires not only retaining all six terms in Eq. \eqref{eq: S_lmw}, rather than keeping just the dominant components in the near-horizon region, but also making no assumptions of slow variance, placing the complete $t$- and $\omega-$ dependence into the complex phase of the integrand.

Including the full source in our steepest-descent approximation of Eq. \eqref{eq: I_lm} introduces two complications: first, its near-horizon expansion contains the screening factors $(\omega-\omega_\text{MB}^{(1)})(\omega-\omega_\text{MB}^{(2)})$ \cite{SizhengPaper1, Zimmerman&Chen}. These zeros coincide with the asymptotic frequencies of the dominant and leading subdominant saddles, introducing logarithmic branch points into the integrand's phase and invalidating the Gaussian saddle approximation.

We address this problem by treating the six terms in Eq. \eqref{eq: S_lmw} separately, for which the screening factors vanish and each phase remains analytic. Defining $\Phi_j(t,\omega)\equiv-i\log[\mathcal{S}_j(t,\omega)]$, with $j\in\{nn0,\bar mn0,\ldots\}$, the saddle points satisfy
\begin{equation}
    \partial_t\Phi_j(t,\omega_s)|_{t_s}=0,\quad \partial_\omega\Phi_j(t_s,\omega)|_{\omega_s}-u=0.
    \label{eq: saddle point conds}
\end{equation}
A near-horizon approximation generalizes the asymptotic saddle frequency in Eq. \eqref{eq: omega_G} to each term, with $u$ still equal to $t-r_*$ but now $\omega_{G,j}=\mathrm{Re}(\omega_G)+N(j)\times\mathrm{Im}(\omega_G)$,
where $N(j)=1$ for $j\in\{\bar m\bar m0,\bar m\bar m1,\bar m\bar m2\}$, $N(j)=2$ for $j\in\{\bar mn0,\bar mn1\}$, and $N(j)=3$ for $j=nn0$. The corresponding horizon-limit frequencies are the first three Matsubara modes. We emphasize that we do not assume the saddle frequencies are equal to $\omega_{G,j}$ -- it merely provides an asymptotic reference.

The second problem is that with the full frequency dependence retained, $\partial_\omega\Phi_j$ is complex for real $t$, yielding complex retarded times $u$. To compute the steepest-descent waveform for real $u$, we must analytically continue the trajectory, source coefficients, and homogeneous solutions to complex $t$, $r$, and $\phi$, seeking saddle pairs $(t_s,\omega_s)$ for which $u$ is real. For each real trajectory time $t$, we solve for $\tau(t)$ such that the saddle time $t_s=t+i\tau(t)$ satisfies both $\partial_t\Phi_j(t_s,\omega_s)=0$ and $\operatorname{Im}[\partial_\omega\Phi_j(t_s,\omega_s)]=0$. Invoking complex coordinate time is not entirely new --- it has been used, for example, to describe bouncing singularities that predict the radii of convergence for the sums which generate the QNM and prompt portions of the Schwarzschild Green's function \cite{Prompt4}, though we will show that the physical intuition is different here. We compute the Teukolsky source functions using the package \verb|GeneralizedSasakiNakamura.jl| \cite{RicoPaper1,RicoPaper2,RicoPaper3,RicoPaper4, JHMethod}, with their analytic continuation implemented using sixth-order Taylor expansions about the real axis.

\ssec{Saddle quantities and waveforms}
\begin{figure}[t]
    \centering
    \includegraphics[width=\linewidth]{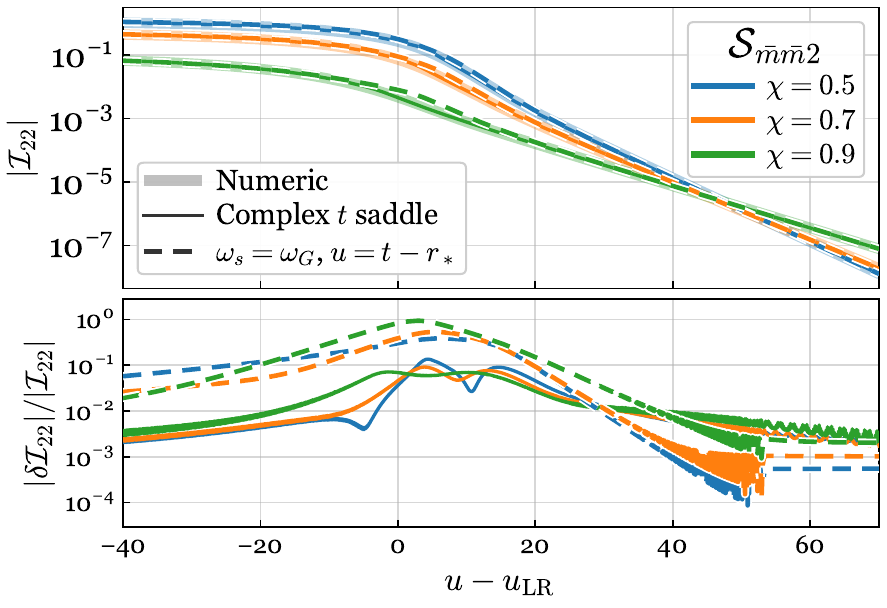}
    \caption{Magnitude of the pure-drive waveform from $\mathcal{S}_{\bar m\bar m2}$, computed using the analytically continued saddle (solid) and with $\omega_{G,j}$ (dashed), for three spins. The lower panel shows fractional differences from direct numerical integration.}
    \label{fig: steepest descent waveforms term 6}
\end{figure}
There is no guarantee that just a single complex saddle exists for each $u$ -- in fact, for some terms, the expected asymptotic frequency at early times (twice the orbital frequency) and late times ($\Omega_{\text{MB}}^{(N(j))}$) lie on separate saddle branches, labeled as ``forward'' and ``backward'', respectively. We find that when two branches emerge, an avoided crossing in $\omega_s$ occurs near $u_\text{LR}$; the waveform must be constructed using the method of \cite{ChesterFriedmanUrsell} in this case. In the vicinity of the light-ring crossing, we find that even when both asymptotic limits lie on the same saddle branch, the stationary frequency $\omega_s(u)$ can deviate noticeably from the simplified model's prediction of $\omega_s=\omega_G$, $u=t-r_*$. For $\mathcal{S}_{nn0}$, we additionally regularize the pole at $\omega_\text{MB}^{(3)}$ in $R_\text{in}$ and restore its contribution at the waveform level. The Supplemental Material presents the complete set of complex saddle quantities for all six terms and three spins.

Figure \ref{fig: steepest descent waveforms term 6} compares the resulting pure-drive waveform from $\mathcal{S}_{\bar m\bar m2}$ with direct numerical integration over the real $t$ and $\omega$ axes and the simplified $\omega_{G,j}$ model. The latter reproduces the early- and late-time behavior but deviates as the waveform turns over, when the stationary frequency evolves rapidly and the $\omega$-dependence of the source prefactors becomes important. The analytically continued saddle agrees with the numerical integration roughly an order of magnitude in fractional error better than the simplified model. Unlike the simpler prescription $\mathcal{I}_{22}(u)\sim\exp(-i\int \omega_G(t(u'))\,du')$ of \cite{SizhengPaper1, SizhengPaper2}, our approach also accounts for the source-phase curvature and requires no fitted rescaling or time shift.

The saddle frequency approaches its horizon-mode limit only after the waveform has fallen well below its peak. This supports the conclusion of \cite{DirectWaveEvolution} that modeling the direct wave with a single mode at the fixed horizon frequency is justified only sufficiently late in the signal; at intermediate times, its rapidly evolving complex frequency must be properly accounted for.

\ssec{Physical correspondence of complex time}
The complex saddle time provides a way to connect the steepest-descent approximation to the emission physics. Unlike the simplified model, which assigns radiation emitted at time $t$ to the observer time $u=t-r_*(t)$ on the light-cone, our method identifies complex saddle times $t_s$ that map onto real $u$. To interpret these times, we return to the response-filtered kernel, broken into its individual source components $W_j\equiv W_{\text{resp},j}$.

\begin{figure}[t]
    \centering
    \includegraphics[width=\linewidth]{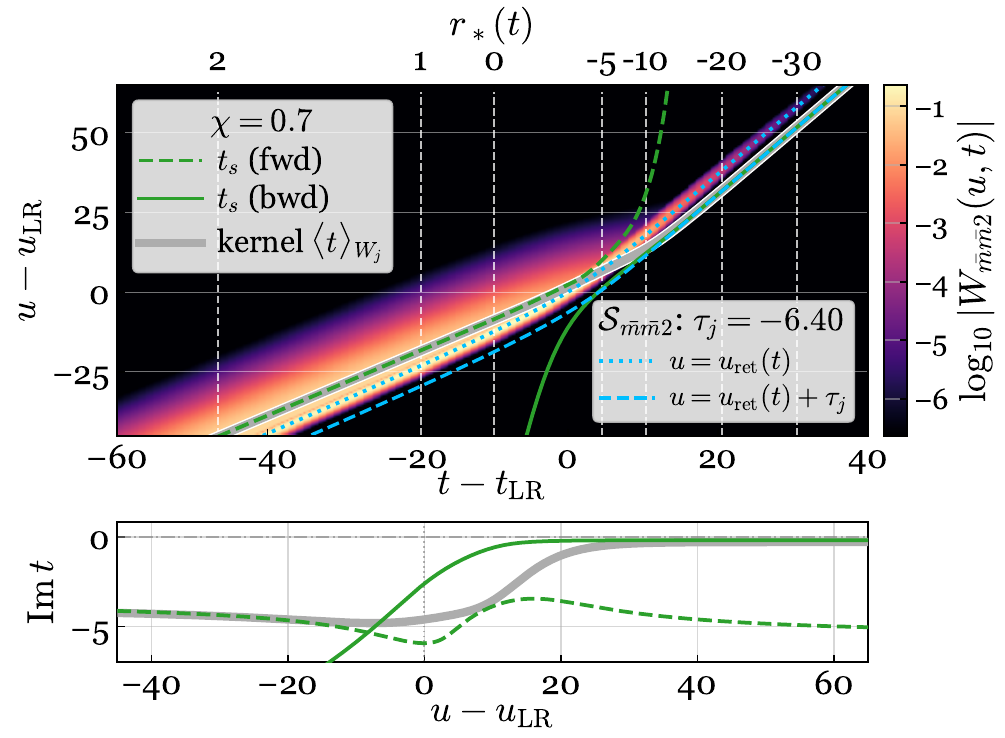}
    \caption{\textit{Top:} The magnitude of the kernel $W_{\bar m\bar m2}(u,t)$, along with the real $\mathrm{Re}[\langle t\rangle_{W_{\bar m\bar m2}}]$, $\mathrm{Re}(t_s)$ for the two complex saddles identified for $\mathcal{S}_{\bar m\bar m2}$, the line $u_\text{ret}=t-r_*$, and the same line plus the expected late time group delay. \textit{Bottom:} The imaginary part of $\langle t\rangle_{W_{\bar m\bar m2}}$ and $t_s$ for the two saddles. In both panels, the saddle quantities match the centroid except in a region around $t_\text{LR}$ where the dominant saddle switches.}
    \label{fig:kernel vs saddle}
\end{figure}

Figure \ref{fig:kernel vs saddle} shows the kernel for $\mathcal{S}_{\bar m\bar m2}$, whose peak lies on the light-cone $u_\text{ret}=t-r_*$. The finite extension into the acausal region arises from the Gaussian frequency filter used to regularize the integral. We characterize the coherent contribution using the complex centroid
\begin{equation}
    \langle t \rangle_{W_j}(u)\equiv\frac{\int dt\,t\,W_j(u,t)}{\int dt\,W_j(u,t)}.
\end{equation}
The real part denotes the time of the coherent contribution to the waveform, while the imaginary part captures the product of the phase variation across the coherent region and its duration. Both parts closely track the corresponding saddle time, except near the light-ring crossing, where the dominant contribution switches between saddle branches. Results for all source terms and spins appear in the Supplemental Material.

At late times, $\mathrm{Re}[\langle t\rangle_{W_j}]$ follows the light-cone with a constant group delay,
\begin{equation}
    \tau_j=\mathrm{Re}[-i\partial_\omega({A_jR_\text{in}^{(j)}}/{R_\text{in}})|_{r\rightarrow r_+,\,\omega\rightarrow\omega_{H,j}}],
\end{equation}
where $R_\text{in}^{(j)}$ denotes the relevant radial derivative in Eq. \eqref{eq: S_lmw}. Although $\tau_j$ can be negative, this does not violate causality: the centroid of a complex kernel need not lie within its support on the real $t$ axis.

\ssec{Phase coherence in the full source}
Near-horizon expansions predict strong cancellations between the three source terms which asymptote to $\omega_\text{MB}^{(1)}$ and between the two source terms which asymptote to $\omega_\text{MB}^{(2)}$, leaving residual contributions decaying at $\omega_\text{MB}^{(3)}$. These cancellations are demonstrated both analytically and numerically in the Supplemental Material and motivate computing the centroid of the summed kernel $W_\text{resp}\equiv\sum_jW_j$ shown in Figure \ref{fig:filtered kernels}.

After the light-ring crossing, the full-source centroid becomes ill-behaved because the screening cancellations nearly eliminate the $n=1$ and $n=2$ contributions to its denominator, leaving the much smaller $n=3$ residual. Their first moments do not cancel in the same way, however, because the additional $t$ weighting changes the numerator. Consequently, the full centroid no longer reliably tracks the relevant emission time.

To recover a physically meaningful centroid, we isolate the $n=3$ component of each kernel slice by fitting the post-peak wake to
\begin{equation}
    W(u,t)\simeq\sum_{n=1}^4 b_n(t)\exp[-i\omega_\text{MB}^{(n)}(u-u_\text{ret})],
\end{equation}
and smoothly extending the fitted $n=3$ component to the light-cone. Removing the non-cancelling $n=1$ and $2$ moments restores a well-behaved centroid, which follows the light-cone at late times with a slight group delay $\sim 3M$. The behavior of the centroid appears in the Supplemental Material alongside the screening cancellations.

\begin{figure}
    \centering
    \includegraphics[width=0.95\linewidth]{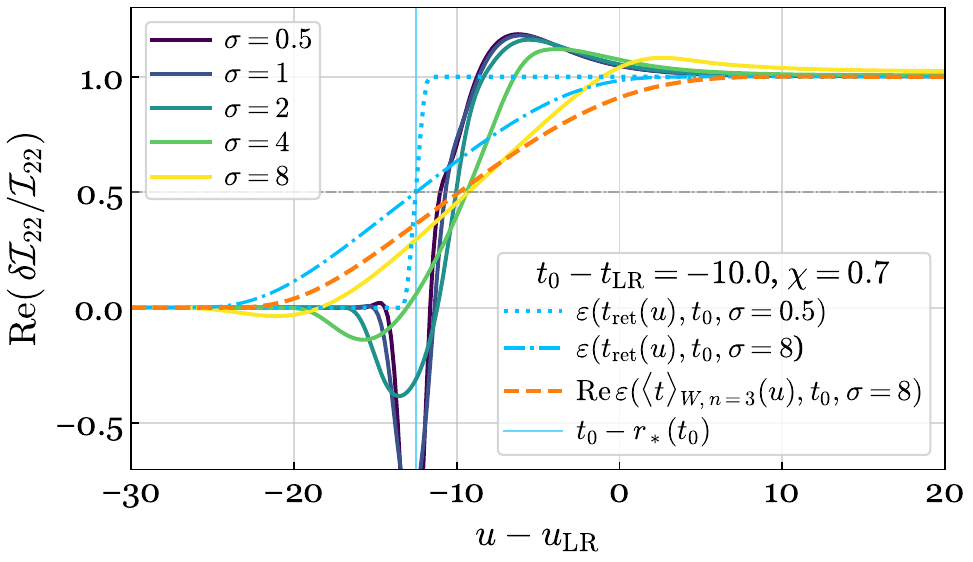}
    \caption{Response of the pure-drive waveform to a smooth source modification $\varepsilon(t)$ that rises from zero to one, centered at $t_0$ with width $\sigma$. Narrow modifications produce a strong response near the light-cone arrival (cyan); broader modifications follow more closely the prediction from the isolated $n=3$ centroid time $\langle t\rangle_{W,n=3}$ (orange).}
    \label{fig:source kernel response}
\end{figure}

The role of the centroid time becomes particularly apparent when considering the imprint of modifications to the source on the pure-drive waveform. In Figure \ref{fig:source kernel response}, we multiply the full source by $1+\varepsilon(t,t_0,\sigma)$, where $\varepsilon(t,t_0,\sigma)$ is a Planck-taper which ramps smoothly from zero to one with $\varepsilon(t_0)=1/2$ and $\dot{\varepsilon}(t_0)=1/(2\sigma)$. We then compute the resulting change in the pure-drive waveform $\delta \mathcal{I}_{22}/\mathcal{I}_{22}$ and compare it to the qualitative prediction $\varepsilon(t(u))$, where $t(u)$ is set by the light-cone time or by the isolated $n=3$ centroid time $\langle t\rangle_{W,n=3}$.

When the ramp's width $\sigma$ is small, the waveform responds sharply near the light-cone arrival $u=t_0-r_*(t_0)$. As the ramp broadens, the response increasingly follows the centroid-time prediction, lagging behind the light-cone arrival prediction by a few $M$. 
Thus, localized source changes register in the observed waveform near their light-cone arrival, whereas modifications extended over the coherent width of the kernel are imprinted as though applied at the delayed centroid time.

\ssec{Discussion}
In this letter, we have shown that the response filter $2i\omega B_\text{in}$ isolates the contribution of individual segments of the plunge to the resulting direct wave. Compared with commonly used rational filters, it produces a more localized kernel mapping emission along the trajectory to the observed direct wave. As an example, we showed that it enhances the \textit{relative} contribution of the resulting direct wave sourced inside the light-ring, typically buried beneath its external counterpart in the unfiltered signal. Localization increases as the particle approaches the horizon, but the source magnitude also decays there, potentially limiting the SNR of the encoded plunge dynamics.

We demonstrated a generalized steepest-descent approximation to the direct wave, obtained via analytic continuation of the Teukolsky homogeneous solutions and source coefficients to complex $t$, $r$, and $\phi$.  Retaining the full frequency dependence of the integrand and treating source terms separately to avoid screening-factor branch points eliminates the need for fitted rescalings or time shifts while improving substantially on the agreement with numerical results obtained by previous approaches.

Reinstating the black-hole response $(2i\omega B_\text{in})^{-1}$ converts the pure-drive waveform's late-time decay at the $n\geq3$ Matsubara modes into high-overtone quasinormal modes through Fano resonances \cite{KerrDynamical}. However, when radiation emitted near the light-ring reaches the observer, the saddle frequency remains far from its asymptotic limit. Thus, the direct wave sourced near and inside the light-ring should not be expected to admit a fixed Matsubara or quasinormal-mode description (depending on the choice of QNM filter). Instead, its complex frequency evolves rapidly on timescales of $\sim 10M$ (see the Supplemental Material), consistent with \cite{DirectWaveEvolution}.

The complex saddle time closely tracks the coordinate-time-weighted centroid of the kernel. Rapid phase variation across the kernel peak on the light-cone produces destructive interference, whereas this centroid identifies where the emission is approximately coherent.
For the full source, cancellations among terms with different asymptotic Matsubara frequencies make the ordinary kernel centroid unreliable. Isolating the residual component decaying at $\omega_\text{MB}^{(3)}$ restores a meaningful centroid and clarifies how source modifications appear in the pure-drive waveform. We observed that a temporally localized modification registers predominantly at its light-cone arrival, whereas an extended modification is imprinted as though it were introduced at the kernel's centroid time, lagging a few $M$ behind the light-cone.

Such temporally-delocalized modifications could arise from accretion disks or dark-matter overdensities \cite{Barausse:2014tra, Kavanagh:2020cfn, Coogan:2021uqv, Cardoso:2021wlq, Cardoso:2022whc}, ultralight boson clouds \cite{Baumann:2018vus, Brito:2023pyl, Duque:2023seg, Li:2025ffh, Li:2026foy}, or deviations from the background metric or theory of gravity \cite{LaHaye:2025ley, Vigeland:2009pr, Collins:2004ex, Maselli:2020zgv, Maselli:2021men, Roy:2025kra}. Our results inform expectations for how alterations acting over extended timescales, particularly near the light-ring, may be encoded in the direct wave. The response-filtered kernel may also be broadly useful for tests of general relativity, either in identifying a modified response function through the extent of localization produced by the GR response filter or a modified source itself; more work is needed to establish this definitively.

Our generalized steepest-descent method could be adapted to rational filters, though the poles from the filters and $B_\text{in}^{-1}$ would require careful handling. Extending the analysis beyond circular, equatorial plunges to inclined or non-circular trajectories would further clarify how source dynamics shape the direct wave and whether it can probe horizon properties beyond a single characteristic asymptotic frequency. Applying the method to effective-one-body trajectories could also yield direct-wave models for response-filtered numerical-relativity waveforms of comparable-mass mergers, further connecting the particle-plunge and comparable-mass regimes.

\ssec{Acknowledgments}
A.L. is grateful for support from the Fannie and John Hertz Foundation through a Hertz Fellowship. A.L., C.W. and Y.C. are supported by the Brinson Foundation, the Simons Foundation (Award Number 568762), and by US NSF Grants PHY--2309211 and PHY--2309231. L.S. acknowledges support by the Australian Research Council (ARC) Centre of Excellence for Gravitational Wave Discovery (OzGrav), Project Number CE230100016, and by the ARC Discovery Early Career Researcher Award, Project Number DE240100206. Research at Perimeter Institute is supported in part by the Government of Canada through the Department of Innovation, Science and Economic Development and by the Province of Ontario through the Ministry of Colleges and Universities. We acknowledge the use of Claude (Anthropic) for assistance with numerical analysis; all results were verified by the authors.

\bibliography{main}

@article{Zimmerman&Chen,
    author = "Zimmerman, Aaron and Chen, Yanbei",
    title = "{New Generic Ringdown Frequencies at the Birth of a Kerr Black Hole}",
    eprint = "1106.0782",
    archivePrefix = "arXiv",
    primaryClass = "gr-qc",
    doi = "10.1103/PhysRevD.84.084012",
    journal = "Phys. Rev. D",
    volume = "84",
    pages = "084012",
    year = "2011"
}

@article{RicoPaper1,
    author = "Lo, Rico K. L.",
    title = "{Recipes for computing radiation from a Kerr black hole using a generalized Sasaki-Nakamura formalism: Homogeneous solutions}",
    eprint = "2306.16469",
    archivePrefix = "arXiv",
    primaryClass = "gr-qc",
    doi = "10.1103/PhysRevD.110.124070",
    journal = "Phys. Rev. D",
    volume = "110",
    number = "12",
    pages = "124070",
    year = "2024"
}

@article{RicoPaper2,
    author = "Lo, Rico K. L. and Sabani, Leart and Cardoso, Vitor",
    title = "{Quasinormal modes and excitation factors of Kerr black holes}",
    eprint = "2504.00084",
    archivePrefix = "arXiv",
    primaryClass = "gr-qc",
    doi = "10.1103/PhysRevD.111.124002",
    journal = "Phys. Rev. D",
    volume = "111",
    number = "12",
    pages = "124002",
    year = "2025"
}

@article{RicoPaper3,
    author = "Yin, Yucheng and Lo, Rico K. L. and Chen, Xian",
    title = "{Gravitational radiation from Kerr black holes using the Sasaki-Nakamura formalism: Waveforms and fluxes at infinity}",
    eprint = "2511.08673",
    archivePrefix = "arXiv",
    primaryClass = "gr-qc",
    doi = "10.1103/9ngz-k1lr",
    journal = "Phys. Rev. D",
    volume = "113",
    number = "12",
    pages = "124007",
    year = "2026"
}

@article{RicoPaper4,
    author = "Lo, Rico K. L. and Yin, Yucheng",
    title = "{Near-horizon gravitational perturbations of rotating black holes}",
    eprint = "2512.07937",
    archivePrefix = "arXiv",
    primaryClass = "gr-qc",
    doi = "10.1103/bljh-l413",
    journal = "Phys. Rev. D",
    volume = "113",
    number = "6",
    pages = "L061505",
    year = "2026"
}

@article{JHMethod,
    author = "Jiang, Ye and Han, Wen-Biao",
    title = "{New high-performing method for solving the homogeneous Teukolsky equation}",
    eprint = "2507.15363",
    archivePrefix = "arXiv",
    primaryClass = "gr-qc",
    doi = "10.1103/kkmt-fbjb",
    journal = "Phys. Rev. D",
    volume = "113",
    number = "2",
    pages = "024001",
    year = "2026"
}

@misc{AdrianPaper1,
    author = "Chung, Adrian Ka-Wai and Lam, Kelvin Ka-Ho and Liu, Anna and Yunes, Nicolas",
    title = "{Measuring a Black Hole's Area Immediately after Merger: A Direct-Wave Test of Hawking's Area Law}",
    eprint = "2606.06592",
    archivePrefix = "arXiv",
    primaryClass = "gr-qc",
    reportNumber = "LIGO PNP P2600278",
    month = "6",
    year = "2026"
}

@misc{AdrianPaper2,
    author = "Dyer, Richard and Chung, Adrian Ka-Wai and Moore, Christopher J.",
    title = "{Modeling Direct Waves in Binary Black Hole Ringdowns}",
    eprint = "2606.25021",
    archivePrefix = "arXiv",
    primaryClass = "gr-qc",
    month = "6",
    year = "2026"
}

@misc{SizhengPaper1,
    author = "Oshita, Naritaka and Ma, Sizheng and Chen, Yanbei and Yang, Huan",
    title = "{Probing Direct Waves in Black Hole Ringdowns}",
    eprint = "2509.09165",
    archivePrefix = "arXiv",
    primaryClass = "gr-qc",
    reportNumber = "YITP-25-141, RIKEN-iTHEMS-Report-25",
    month = "9",
    year = "2025"
}

@article{SizhengPaper2,
    author = "Lu, Neil and Ma, Sizheng and Piccinni, Ornella J. and Chen, Yanbei and Sun, Ling",
    title = "{GW250114 reveals signatures of post-merger black-hole horizon}",
    eprint = "2510.01001",
    archivePrefix = "arXiv",
    primaryClass = "gr-qc",
    doi = "10.1038/s41586-026-10696-0",
    journal = "Nature",
    volume = "655",
    number = "8121",
    pages = "71--75",
    year = "2026"
}

@misc{SizhengPaper3,
    author = "Ma, Sizheng and Wang, Hai-Yang",
    title = "{Foundations of Direct Waves in Schwarzschild Ringdown}",
    eprint = "2607.06677",
    archivePrefix = "arXiv",
    primaryClass = "gr-qc",
    month = "7",
    year = "2026"
}

@misc{KerrDynamical,
    author = "Weller, Colin and Laeuger, Andrew and Li, Fulin and Lu, Neil and Nathan, Rowina S. and Ma, Sizheng and Chen, Yanbei and Sun, Ling",
    title = "{From driven oscillations to free ringdown: a particle plunging into Kerr}",
    eprint = "2609.09531",
    archivePrefix = "arXiv",
    primaryClass = "gr-qc",
    month = "9",
    year = "2026"
}

@misc{DirectWaveEvolution,
    author = "Sun, Ling and Ma, Sizheng and Chen, Patrick P. and Laeuger, Andrew and Lu, Neil and Chen, Yanbei",
    title = "{Complex frequency evolution of direct waves from binary black hole mergers}",
    eprint = "2608.23209",
    archivePrefix = "arXiv",
    primaryClass = "gr-qc",
    month = "8",
    year = "2026"
}

@misc{Nakamoto:2026rah,
    author = "Nakamoto, Nao and Oshita, Naritaka and Takeda, Hiroki",
    title = "{Theoretical Aspects of Direct Waves in Kerr Black Holes: Pole-Splitting Method for Ringdown Analysis}",
    eprint = "2609.11908",
    archivePrefix = "arXiv",
    primaryClass = "gr-qc",
    reportNumber = "RIKEN-iTHEMS-Report-26",
    month = "9",
    year = "2026"
}

@misc{Kubota:2026qjh,
    author = "Kubota, Kei-ichiro and Manita, Yusuke and Motohashi, Hayato and Watarai, Daiki",
    title = "{Exact Cancellation of Horizon Modes in Kerr Ringdown}",
    eprint = "2609.07326",
    archivePrefix = "arXiv",
    primaryClass = "gr-qc",
    reportNumber = "RESCEU-23/26",
    month = "9",
    year = "2026"
}

@article{MinoReview,
    author = "Mino, Yasushi and Sasaki, Misao and Shibata, Masaru and Tagoshi, Hideyuki and Tanaka, Takahiro",
    title = "{Black hole perturbation: Chapter 1}",
    eprint = "gr-qc/9712057",
    archivePrefix = "arXiv",
    doi = "10.1143/PTPS.128.1",
    journal = "Prog. Theor. Phys. Suppl.",
    volume = "128",
    pages = "1--121",
    year = "1997"
}

@article{SasakiReview,
    author = "Sasaki, Misao and Tagoshi, Hideyuki",
    title = "{Analytic black hole perturbation approach to gravitational radiation}",
    eprint = "gr-qc/0306120",
    archivePrefix = "arXiv",
    doi = "10.12942/lrr-2003-6",
    journal = "Living Rev. Rel.",
    volume = "6",
    pages = "6",
    year = "2003"
}

@misc{KankaniPaper1,
    author = "Kankani, Anuj and McWilliams, Sean T.",
    title = "{Direct Waves in Black-Hole Binary Mergers: Insights from the Backwards One Body Model}",
    eprint = "2603.15474",
    archivePrefix = "arXiv",
    primaryClass = "gr-qc",
    month = "3",
    year = "2026"
}

@misc{KankaniPaper2,
    author = "Kankani, Anuj and McWilliams, Sean T.",
    title = "{The Direct Wave is Not a Meaningful Test of Horizon Properties}",
    eprint = "2607.02380",
    archivePrefix = "arXiv",
    primaryClass = "gr-qc",
    month = "7",
    year = "2026"
}

@article{Marina1,
    author = "De Amicis, Marina and Cannizzaro, Enrico and Carullo, Gregorio and Sberna, Laura",
    title = "{Dynamical quasinormal mode excitation}",
    eprint = "2506.21668",
    archivePrefix = "arXiv",
    primaryClass = "gr-qc",
    doi = "10.1103/bgjv-lvyd",
    journal = "Phys. Rev. D",
    volume = "113",
    number = "2",
    pages = "024048",
    year = "2026"
}

@misc{Marina2,
    author = "De Amicis, Marina and Cannizzaro, Enrico and Carullo, Gregorio and Kuntz, Adrien and Sberna, Laura",
    title = "{Dynamical quasinormal mode excitation II: propagation and convergence in Schwarzschild}",
    eprint = "2605.16492",
    archivePrefix = "arXiv",
    primaryClass = "gr-qc",
    month = "5",
    year = "2026",
    journal = {}
}

@article{Prompt1,
    author = "Su, Junquan and Khera, Neev and Casals, Marc and Ma, Sizheng and Chowdhuri, Abhishek and Yang, Huan",
    title = "{Decomposition of the Schwarzschild Green{\textquoteright}s function}",
    eprint = "2601.22015",
    archivePrefix = "arXiv",
    primaryClass = "gr-qc",
    doi = "10.1103/3n8r-8kbb",
    journal = "Phys. Rev. D",
    volume = "113",
    number = "10",
    pages = "104013",
    year = "2026"
}

@misc{Prompt2,
    author = "Ma, Sizheng",
    title = "{Prompt Response from Plunging Sources in Schwarzschild Spacetime}",
    eprint = "2604.08680",
    archivePrefix = "arXiv",
    primaryClass = "gr-qc",
    month = "4",
    year = "2026"
}

@misc{Prompt3,
    author = "Arnaudo, Paolo and Carballo, Javier and Withers, Benjamin",
    title = "{Beyond quasinormal modes: a complete mode decomposition of black hole perturbations}",
    eprint = "2510.18956",
    archivePrefix = "arXiv",
    primaryClass = "gr-qc",
    month = "10",
    year = "2025"
}

@misc{Prompt4,
    author = "Arnaudo, Paolo and Withers, Benjamin",
    title = "{Bouncing singularities in Schwarzschild: a geometric origin of the QNM convergence region}",
    eprint = "2605.16489",
    archivePrefix = "arXiv",
    primaryClass = "gr-qc",
    month = "5",
    year = "2026"
}

@article{QQNM1,
    author = "Bourg, Patrick and Panosso Macedo, Rodrigo and Spiers, Andrew and Leather, Benjamin and Bonga, B{\'e}atrice and Pound, Adam",
    title = "{Quadratic Quasinormal Mode Dependence on Linear Mode Parity}",
    eprint = "2405.10270",
    archivePrefix = "arXiv",
    primaryClass = "gr-qc",
    doi = "10.1103/PhysRevLett.134.061401",
    journal = "Phys. Rev. Lett.",
    volume = "134",
    number = "6",
    pages = "061401",
    year = "2025"
}

@article{QQNM2,
    author = "Bourg, Patrick and Panosso Macedo, Rodrigo and Spiers, Andrew and Leather, Benjamin and B{\'e}atrice, Bonga and Pound, Adam",
    title = "{Quadratic quasinormal modes at null infinity on a Schwarzschild spacetime}",
    eprint = "2503.07432",
    archivePrefix = "arXiv",
    primaryClass = "gr-qc",
    doi = "10.1103/fbz4-qsvn",
    journal = "Phys. Rev. D",
    volume = "112",
    number = "4",
    pages = "044049",
    year = "2025"
}

@article{QQNM3,
    author = "Bucciotti, Bruno and Juliano, Leonardo and Kuntz, Adrien and Trincherini, Enrico",
    title = "{Quadratic quasinormal modes of a Schwarzschild black hole}",
    eprint = "2405.06012",
    archivePrefix = "arXiv",
    primaryClass = "gr-qc",
    doi = "10.1103/PhysRevD.110.104048",
    journal = "Phys. Rev. D",
    volume = "110",
    number = "10",
    pages = "104048",
    year = "2024"
}

@article{QQNM4,
    author = "Ma, Sizheng and Yang, Huan",
    title = "{Excitation of quadratic quasinormal modes for Kerr black holes}",
    eprint = "2401.15516",
    archivePrefix = "arXiv",
    primaryClass = "gr-qc",
    doi = "10.1103/PhysRevD.109.104070",
    journal = "Phys. Rev. D",
    volume = "109",
    number = "10",
    pages = "104070",
    year = "2024"
}

@article{QQNM5,
    author = "Khera, Neev and Ma, Sizheng and Yang, Huan",
    title = "{Quadratic Mode Couplings in Rotating Black Holes and Their Detectability}",
    eprint = "2410.14529",
    archivePrefix = "arXiv",
    primaryClass = "gr-qc",
    doi = "10.1103/PhysRevLett.134.211404",
    journal = "Phys. Rev. Lett.",
    volume = "134",
    number = "21",
    pages = "211404",
    year = "2025"
}

@article{QQNM6,
    author = "Mitman, Keefe and others",
    title = "{Nonlinearities in Black Hole Ringdowns}",
    eprint = "2208.07380",
    archivePrefix = "arXiv",
    primaryClass = "gr-qc",
    doi = "10.1103/PhysRevLett.130.081402",
    journal = "Phys. Rev. Lett.",
    volume = "130",
    number = "8",
    pages = "081402",
    year = "2023"
}

@article{QQNM7,
    author = "Mitman, Keefe and others",
    title = "{Probing the ringdown perturbation in binary black hole coalescences with an improved quasinormal mode extraction algorithm}",
    eprint = "2503.09678",
    archivePrefix = "arXiv",
    primaryClass = "gr-qc",
    doi = "10.1103/qq1g-jlnw",
    journal = "Phys. Rev. D",
    volume = "112",
    number = "6",
    pages = "064016",
    year = "2025"
}

@article{QQNM8,
    author = "Zhu, Hengrui and others",
    title = "{Nonlinear effects in black hole ringdown from scattering experiments: Spin and initial data dependence of quadratic mode coupling}",
    eprint = "2401.00805",
    archivePrefix = "arXiv",
    primaryClass = "gr-qc",
    doi = "10.1103/PhysRevD.109.104050",
    journal = "Phys. Rev. D",
    volume = "109",
    number = "10",
    pages = "104050",
    year = "2024"
}

@article{QNMs1,
    author = "Vishveshwara, C. V.",
    title = "{Scattering of Gravitational Radiation by a Schwarzschild Black-hole}",
    doi = "10.1038/227936a0",
    journal = "Nature",
    volume = "227",
    pages = "936--938",
    year = "1970"
}

@article{QNMs2,
    author = "Press, William H.",
    title = "{Long Wave Trains of Gravitational Waves from a Vibrating Black Hole}",
    reportNumber = "OAP-262",
    doi = "10.1086/180849",
    journal = "Astrophys. J. Lett.",
    volume = "170",
    pages = "L105--L108",
    year = "1971"
}

@article{QNMs3,
    author = "Davis, M. and Ruffini, R. and Press, W. H. and Price, R. H.",
    title = "{Gravitational radiation from a particle falling radially into a Schwarzschild black hole}",
    doi = "10.1103/PhysRevLett.27.1466",
    journal = "Phys. Rev. Lett.",
    volume = "27",
    pages = "1466--1469",
    year = "1971"
}

@article{QNMs4,
    author = "Leaver, E. W.",
    title = "{An Analytic representation for the quasi normal modes of Kerr black holes}",
    doi = "10.1098/rspa.1985.0119",
    journal = "Proc. Roy. Soc. Lond. A",
    volume = "402",
    pages = "285--298",
    year = "1985"
}

@article{QNMs5,
    author = "Andersson, N.",
    title = "{Excitation of Schwarzschild black hole quasinormal modes}",
    doi = "10.1103/PhysRevD.51.353",
    journal = "Phys. Rev. D",
    volume = "51",
    pages = "353--363",
    year = "1995"
}

@article{Tails1,
    author = "Price, Richard H.",
    title = "{Nonspherical perturbations of relativistic gravitational collapse. 1. Scalar and gravitational perturbations}",
    doi = "10.1103/PhysRevD.5.2419",
    journal = "Phys. Rev. D",
    volume = "5",
    pages = "2419--2438",
    year = "1972"
}

@article{Tails2,
    author = "Leaver, Edward W.",
    title = "{Spectral decomposition of the perturbation response of the Schwarzschild geometry}",
    doi = "10.1103/PhysRevD.34.384",
    journal = "Phys. Rev. D",
    volume = "34",
    pages = "384--408",
    year = "1986"
}

@article{Tails3,
    author = "Barack, Leor",
    title = "{Late time decay of scalar, electromagnetic, and gravitational perturbations outside rotating black holes}",
    eprint = "gr-qc/9908005",
    archivePrefix = "arXiv",
    doi = "10.1103/PhysRevD.61.024026",
    journal = "Phys. Rev. D",
    volume = "61",
    pages = "024026",
    year = "2000"
}

@article{ChesterFriedmanUrsell,
    title={An extension of the method of steepest descents}, 
    volume={53}, 
    DOI={10.1017/S0305004100032655}, number={3}, 
    journal={Mathematical Proceedings of the Cambridge Philosophical Society}, author={Chester, C. and Friedman, B. and Ursell, F.}, 
    year={1957}, 
    pages={599--611}
    }

@article{MaFilters,
    author = "Ma, Sizheng and Mitman, Keefe and Sun, Ling and Deppe, Nils and H{\'e}bert, Fran{\c{c}}ois and Kidder, Lawrence E. and Moxon, Jordan and Throwe, William and Vu, Nils L. and Chen, Yanbei",
    title = "{Quasinormal-mode filters: A new approach to analyze the gravitational-wave ringdown of binary black-hole mergers}",
    eprint = "2207.10870",
    archivePrefix = "arXiv",
    primaryClass = "gr-qc",
    doi = "10.1103/PhysRevD.106.084036",
    journal = "Phys. Rev. D",
    volume = "106",
    number = "8",
    pages = "084036",
    year = "2022"
}

@article{Li:2026foy,
    author = "Li, Dongjun and Kuan, Hao-Jui and Weller, Colin and Yang, Chaoyi and Lam, Kelvin Ka-Ho and Bourg, Patrick and Chung, Adrian Ka-Wai and Chen, Yanbei and Yunes, Nicol{\'a}s and Yang, Huan",
    title = "{Modeling extreme mass-ratio inspirals in astrophysical and dark-matter environments}",
    eprint = "2610.22480",
    archivePrefix = "arXiv",
    primaryClass = "gr-qc",
    month = "10",
    year = "2026",
    journal = {}
}

@article{LaHaye:2025ley,
    author = "LaHaye, Michael and Weller, Colin and Li, Dongjun and Bourg, Patrick and Chen, Yanbei and Yang, Huan",
    title = "{Evolving extreme mass-ratio inspirals in a perturbed Schwarzschild spacetime}",
    eprint = "2510.16102",
    archivePrefix = "arXiv",
    primaryClass = "gr-qc",
    doi = "10.1103/dy1z-36w5",
    journal = "Phys. Rev. D",
    volume = "113",
    number = "2",
    pages = "024069",
    year = "2026"
}

@article{Li:2025ffh,
    author = "Li, Dongjun and Weller, Colin and Bourg, Patrick and LaHaye, Michael and Yunes, Nicol{\'a}s and Yang, Huan",
    title = "{Extreme mass-ratio inspiral within an ultralight scalar cloud: Scalar radiation}",
    eprint = "2507.02045",
    archivePrefix = "arXiv",
    primaryClass = "gr-qc",
    doi = "10.1103/7l9s-g21j",
    journal = "Phys. Rev. D",
    volume = "112",
    number = "8",
    pages = "084057",
    year = "2025"
}

@article{Duque:2023seg,
    author = "Duque, Francisco and Macedo, Caio F. B. and Vicente, Rodrigo and Cardoso, Vitor",
    title = "{Extreme-Mass-Ratio Inspirals in Ultralight Dark Matter}",
    eprint = "2312.06767",
    archivePrefix = "arXiv",
    primaryClass = "gr-qc",
    doi = "10.1103/PhysRevLett.133.121404",
    journal = "Phys. Rev. Lett.",
    volume = "133",
    number = "12",
    pages = "121404",
    year = "2024"
}

@article{Brito:2023pyl,
    author = "Brito, Richard and Shah, Shreya",
    title = "{Extreme mass-ratio inspirals into black holes surrounded by scalar clouds}",
    eprint = "2307.16093",
    archivePrefix = "arXiv",
    primaryClass = "gr-qc",
    doi = "10.1103/PhysRevD.108.084019",
    journal = "Phys. Rev. D",
    volume = "108",
    number = "8",
    pages = "084019",
    year = "2023",
    note = "[Erratum: Phys.Rev.D 110, 109902 (2024)]"
}

@article{Baumann:2018vus,
    author = "Baumann, Daniel and Chia, Horng Sheng and Porto, Rafael A.",
    title = "{Probing Ultralight Bosons with Binary Black Holes}",
    eprint = "1804.03208",
    archivePrefix = "arXiv",
    primaryClass = "gr-qc",
    reportNumber = "DESY-18-060, DESY 18-060",
    doi = "10.1103/PhysRevD.99.044001",
    journal = "Phys. Rev. D",
    volume = "99",
    number = "4",
    pages = "044001",
    year = "2019"
}

@article{Barausse:2014tra,
    author = "Barausse, Enrico and Cardoso, Vitor and Pani, Paolo",
    title = "{Can environmental effects spoil precision gravitational-wave astrophysics?}",
    eprint = "1404.7149",
    archivePrefix = "arXiv",
    primaryClass = "gr-qc",
    doi = "10.1103/PhysRevD.89.104059",
    journal = "Phys. Rev. D",
    volume = "89",
    number = "10",
    pages = "104059",
    year = "2014"
}

@article{Kavanagh:2020cfn,
    author = "Kavanagh, Bradley J. and Nichols, David A. and Bertone, Gianfranco and Gaggero, Daniele",
    title = "{Detecting dark matter around black holes with gravitational waves: Effects of dark-matter dynamics on the gravitational waveform}",
    eprint = "2002.12811",
    archivePrefix = "arXiv",
    primaryClass = "gr-qc",
    doi = "10.1103/PhysRevD.102.083006",
    journal = "Phys. Rev. D",
    volume = "102",
    number = "8",
    pages = "083006",
    year = "2020"
}

@article{Coogan:2021uqv,
    author = "Coogan, Adam and Bertone, Gianfranco and Gaggero, Daniele and Kavanagh, Bradley J. and Nichols, David A.",
    title = "{Measuring the dark matter environments of black hole binaries with gravitational waves}",
    eprint = "2108.04154",
    archivePrefix = "arXiv",
    primaryClass = "gr-qc",
    doi = "10.1103/PhysRevD.105.043009",
    journal = "Phys. Rev. D",
    volume = "105",
    number = "4",
    pages = "043009",
    year = "2022"
}

@article{Cardoso:2021wlq,
    author = "Cardoso, Vitor and Destounis, Kyriakos and Duque, Francisco and Macedo, Rodrigo Panosso and Maselli, Andrea",
    title = "{Black holes in galaxies: Environmental impact on gravitational-wave generation and propagation}",
    eprint = "2109.00005",
    archivePrefix = "arXiv",
    primaryClass = "gr-qc",
    doi = "10.1103/PhysRevD.105.L061501",
    journal = "Phys. Rev. D",
    volume = "105",
    number = "6",
    pages = "L061501",
    year = "2022"
}

@article{Cardoso:2022whc,
    author = "Cardoso, Vitor and Destounis, Kyriakos and Duque, Francisco and Panosso Macedo, Rodrigo and Maselli, Andrea",
    title = "{Gravitational Waves from Extreme-Mass-Ratio Systems in Astrophysical Environments}",
    eprint = "2210.01133",
    archivePrefix = "arXiv",
    primaryClass = "gr-qc",
    doi = "10.1103/PhysRevLett.129.241103",
    journal = "Phys. Rev. Lett.",
    volume = "129",
    number = "24",
    pages = "241103",
    year = "2022"
}

@article{Vigeland:2009pr,
    author = "Vigeland, Sarah J. and Hughes, Scott A.",
    title = "{Spacetime and orbits of bumpy black holes}",
    eprint = "0911.1756",
    archivePrefix = "arXiv",
    primaryClass = "gr-qc",
    doi = "10.1103/PhysRevD.81.024030",
    journal = "Phys. Rev. D",
    volume = "81",
    pages = "024030",
    year = "2010"
}

@article{Collins:2004ex,
    author = "Collins, Nathan A. and Hughes, Scott A.",
    title = "{Towards a formalism for mapping the space-times of massive compact objects: Bumpy black holes and their orbits}",
    eprint = "gr-qc/0402063",
    archivePrefix = "arXiv",
    doi = "10.1103/PhysRevD.69.124022",
    journal = "Phys. Rev. D",
    volume = "69",
    pages = "124022",
    year = "2004"
}

@article{Maselli:2021men,
    author = "Maselli, Andrea and Franchini, Nicola and Gualtieri, Leonardo and Sotiriou, Thomas P. and Barsanti, Susanna and Pani, Paolo",
    title = "{Detecting fundamental fields with LISA observations of gravitational waves from extreme mass-ratio inspirals}",
    eprint = "2106.11325",
    archivePrefix = "arXiv",
    primaryClass = "gr-qc",
    doi = "10.1038/s41550-021-01589-5",
    journal = "Nature Astron.",
    volume = "6",
    number = "4",
    pages = "464--470",
    year = "2022"
}

@article{Maselli:2020zgv,
    author = "Maselli, Andrea and Franchini, Nicola and Gualtieri, Leonardo and Sotiriou, Thomas P.",
    title = "{Detecting scalar fields with Extreme Mass Ratio Inspirals}",
    eprint = "2004.11895",
    archivePrefix = "arXiv",
    primaryClass = "gr-qc",
    doi = "10.1103/PhysRevLett.125.141101",
    journal = "Phys. Rev. Lett.",
    volume = "125",
    number = "14",
    pages = "141101",
    year = "2020"
}

@article{Roy:2025kra,
    author = {Roy, Ayush and K{\"u}chler, Lorenzo and Pound, Adam and Panosso Macedo, Rodrigo},
    title = "{Black Hole Mergers Beyond General Relativity: A Self-Force Approach}",
    eprint = "2510.11793",
    archivePrefix = "arXiv",
    primaryClass = "gr-qc",
    doi = "10.1103/7zjr-l5p8",
    journal = "Phys. Rev. Lett.",
    volume = "136",
    number = "25",
    pages = "251404",
    year = "2026"
}

@article{Dyer:2026sox,
    author = "Dyer, Richard and Kankani, Anuj and Marks, Gareth Arturo and McWilliams, Sean T. and Moore, Christopher J.",
    title = "{Moving Beyond Quasinormal Modes and Reinterpreting Direct Waves in Binary Black Hole Merger-Ringdowns}",
    eprint = "2610.25176",
    archivePrefix = "arXiv",
    primaryClass = "gr-qc",
    month = "10",
    year = "2026",
    journal = {}
}

@article{Kuntz:2026xep,
    author = "Kuntz, Adrien and Della Rocca, Matteo",
    title = "{Vanishing of all redshift modes in Schwarzschild ringdown}",
    eprint = "2606.07173",
    archivePrefix = "arXiv",
    primaryClass = "gr-qc",
    month = "6",
    year = "2026",
    journal = {}
}

@article{Shi2026,
    author = "Jun-Xi Shi and Zhen-Tao He and Tingyuan Jiang and Yu Tian and Dan-Dan Yuan and Hongbao Zhang",
    title = "{Modeling QNMs-Filtered Gravitational Waveforms: Approximation Limits and Filtering Effects}",
    eprint = "2610.30031",
    archivePrefix = "arXiv",
    primaryClass = "gr-qc",
    month = "10",
    year = "2026",
    journal = {}
}

@article{Cheung:2026gfd,
    author = "Cheung, Mark Ho-Yeuk",
    title = "{How much of the black hole ringdown is sourced within the light ring?}",
    eprint = "2608.29466",
    archivePrefix = "arXiv",
    primaryClass = "gr-qc",
    month = "8",
    year = "2026",
    journal = {}
}

\onecolumngrid
\newpage
\appendix

\section*{Supplemental Material}
\subsection*{Deriving the horizon mode screening factors}
Here, we derive the cancellations between source terms which asymptotically approach $\omega_\text{MB}^{(1)}$ and $\omega_\text{MB}^{(2)}$, generating screening factors which leave behind a residual at $\omega_\text{MB}^{(3)}$. The source coefficients $A_j$ are \cite{MinoReview, SasakiReview}
\begin{subequations}
\begin{gather}
    \Apair{A_{nn0}=\frac{-2}{\sqrt{2\pi}\Delta^2}C_{nn}\rho^{-2}\bar\rho^{-1}L_1^\dagger\{\rho^{-4}L_2^\dagger(\rho^3 S)\},}
          {A_{\bar m n1}=\frac{2}{\sqrt{\pi}\Delta}\rho^{-3}C_{\bar mn}[L_2^\dagger S+ia\sin\theta(\bar\rho-\rho)S],}\\
    \Apair{A_{\bar m\bar m0}=-\frac{1}{\sqrt{2\pi}}\rho^{-3}\bar \rho C_{\bar m\bar m}S\Bigl[-i\partial_r(K/\Delta)-\frac{K^2}{\Delta^2}+2i\rho\frac K\Delta\Bigr],}
          {A_{\bar m\bar m1}=-\frac{2}{\sqrt{2\pi}}\rho^{-3}\bar\rho C_{\bar m\bar m}S\Bigl(i\frac K\Delta+\rho\Bigr),}\\
    \Apair{A_{\bar mn0}=\frac{2}{\sqrt{\pi}\Delta}C_{\bar m n}\rho^{-3}\Bigl[(L_2^\dagger S)\Bigl(\frac{iK}{\Delta}+\rho+\bar \rho\Bigr)-a\sin\theta \frac{SK}{\Delta}(\bar\rho-\rho)\Bigr],}
          {A_{\bar m\bar m2}=-\frac{1}{\sqrt{2\pi}}\rho^{-3}\bar \rho C_{\bar m\bar m}S,}
\end{gather}
\end{subequations}
where $\Delta\equiv r^2-2r+a^2$, $\rho\equiv 1/(r-ia\cos\theta)$, $K\equiv (r^2+a^2)\omega-ma$, $S\equiv ~_{-2}S_{\ell m}(a\omega;\theta)$ is the spin-weighted spheroidal harmonic, $L_s^\dagger\equiv \partial_\theta-\frac{m}{\sin\theta}+a\omega\sin\theta+s\cot\theta$, and the $C_j$ coefficients are defined by
\begin{subequations}
\begin{align}
    C_{nn}&=\frac{1}{4\Sigma^3(dt/d\tau)}\Bigl[E(r^2+a^2)-al_z+\Sigma\frac{dr}{d\tau}\Bigr]^2,\\
    C_{\bar mn}&=-\frac{\rho}{2\sqrt{2}\Sigma^2(dt/d\tau)}\Bigl[E(r^2+a^2)-al_z+\Sigma\frac{dr}{d\tau}\Bigr]\Bigl[i\sin\theta\Bigl(aE-\frac{l_z}{\sin^2\theta}\Bigr)+\Sigma\frac{d\theta}{d\tau}\Bigr],\\
    C_{\bar m\bar m}&=\frac{\rho^2}{2\Sigma(dt/d\tau)}\Bigl[i\sin\theta\Bigl(aE-\frac{l_z}{\sin^2\theta}\Bigr)+\Sigma\frac{d\theta}{d\tau}\Bigr]^2,
\end{align}
\end{subequations}
where $\tau$ here is the proper time of the plunging point object and $\Sigma\equiv r^2+a^2\cos^2\theta$. For a massive particle with energy $E$, angular momentum $l_z$, and Carter constant $Q$ ($=0$ for an equatorial orbit), the geodesic equations of motion are
\begin{subequations}
\begin{align}
    \Sigma\frac{d\theta}{d\tau}&=\pm\Bigl[Q-\cos^2\theta\Bigl\{a^2(1-E^2)+\frac{l_z^2}{\sin^2\theta}\Bigr\}\Bigr]^{1/2},\\
    \Sigma\frac{d\varphi}{d\tau}&=-\Bigl(aE-\frac{l_z}{\sin^2\theta}\Bigr)+\frac{a}{\Delta}\Bigl(E(r^2+a^2)-al_z\Bigr),\\
    \Sigma\frac{dt}{d\tau}&=-\Bigl(aE-\frac{l_z}{\sin^2\theta}\Bigr)a\sin^2\theta+\frac{r^2+a^2}{\Delta}\Bigl(E(r^2+a^2)-al_z\Bigr),\\
    \Sigma\frac{dr}{d\tau}&=\pm\Bigl[\{E(r^2+a^2)-al_z\}^2-\Delta\{(aE-l_z)^2+r^2+Q\}\Bigr]^{1/2}.
\end{align}
\end{subequations}

It is useful to extract the leading-order behavior of the homogeneous Teukolsky solutions near the horizon:
\begin{subequations}
\begin{align}
    R_\text{in}(r,\omega)&\approx\Delta^2\exp(-i(\omega-m\Omega_H)r_*),\\
    R'_\text{in}(r,\omega)&\approx\Delta\Bigl(4(r-1)-i(r^2+a^2)(\omega-m\Omega_H)\Bigr)\exp(-i(\omega-m\Omega_H)r_*),\\
    R''_\text{in}(r,\omega)&\approx\Bigl(8(r-1)^2+4\Delta-2ir\Delta(\omega-m\Omega_H)-6i(r-1)(r^2+a^2)(\omega-m\Omega_H)\nonumber\\
    &\qquad-(r^2+a^2)^2(\omega-m\Omega_H)^2\Bigr)\exp(-i(\omega-m\Omega_H)r_*).
\end{align}
\end{subequations}
By inserting these approximations into Eq. \eqref{eq: S_lmw}, one can show that, specializing to the equatorial case $\theta=\pi/2$,
\begin{multline}
    A_{\bar m\bar m0}R_\text{in}-A_{\bar m\bar m1}R'_\text{in}+A_{\bar m\bar m2}R''_\text{in}\\
    \approx -(\omega-\omega_\text{MB}^{(1)})(\omega-\omega_\text{MB}^{(2)})e^{-i(\omega-m\Omega_H)r_*}e^{2\kappa_Hr_*}\times\frac{16(1-a^2)^{1/r_+}e^{-\sqrt{1-a^2}}(l_z-aE)^2(2aE-r_+l_z)S(\pi/2)}{a(4E^2+l_z^2)-4El_z},
\end{multline}
which clearly vanishes at both $\omega=\omega_\text{MB}^{(1)}$ and $\omega=\omega_\text{MB}^{(2)}$. The same set of approximations also give
\begin{multline}
    A_{\bar mn0}R_\text{in}-A_{\bar mn1}R'_\text{in}
    \\
    \approx(\omega-\omega_\text{MB}^{(2)})e^{-i(\omega-m\Omega_H)r_*}e^{4\kappa_Hr_*}\times\frac{16(1-a^2)^{2/r_+}e^{-2\sqrt{1-a^2}}(aE-l_z)(2r_++a^2(E^2-1)-2aEl_z+l_z^2)T(\pi/2)}{(al_z-2r_+E)^2},
\end{multline}
where $T(\theta)\equiv S'(\theta)+(a\omega-m)S(\theta)$ with the prime here denoting $\partial_\theta$ -- this sum evidently vanishes at $\omega=\omega_\text{MB}^{(2)}$. We have thus explicitly produced the screening factors which cause the sum of $\mathcal{I}_{\bar m\bar m0}$, $\mathcal{I}_{\bar m\bar m1}$, and $\mathcal{I}_{\bar m\bar m2}$, each of which decays on its own at $\omega_\text{MB}^{(1)}$, to decay at $\omega_\text{MB}^{(3)}$ in the near-horizon region, and the sum of $\mathcal{I}_{\bar mn0}$ and $\mathcal{I}_{\bar mn1}$, each of which decays at $\omega_\text{MB}^{(2)}$, to also decay at $\omega_\text{MB}^{(3)}$ as the particle nears the horizon, matching the decay of $\mathcal{I}_{nn0}$ and ensuring the complete pure-drive waveform also decays at this rate at late times.

\subsection*{Deriving the horizon mode poles of $R_\text{in}$}
\vspace{-0.5\baselineskip}
Here, we show that $R_\text{in}$ has simple poles at $\omega_\text{MB}^{(N)}=m\Omega_H-iN\kappa_H$ only for $N\geq3$, so that $\omega_\text{MB}^{(3)}$ is the first horizon mode at which the homogeneous solution itself is singular. For $s=-2$, the radial Teukolsky equation reads
\begin{equation}
    \Delta^2\frac{d}{dr}\Bigl(\frac{1}{\Delta}\frac{dR}{dr}\Bigr)+\Bigl[\frac{K^2+4i(r-1)K}{\Delta}-8i\omega r-\lambda\Bigr]R=0,
\end{equation}
where $\lambda$ is the separation constant. Writing $R\propto(r-r_+)^\nu$ near the horizon and defining
\begin{equation}
    p\equiv\frac{(r_+^2+a^2)(\omega-m\Omega_H)}{r_+-r_-}=\frac{\omega-m\Omega_H}{2\kappa_H},
\end{equation}
the indicial equation used in the method of Frobenius at $r=r_+$ factors as
\begin{equation}
    P(\nu)\equiv(\nu-ip)(\nu-2+ip)=0,
\end{equation}
with roots $\nu_\text{in}=2-ip$ and $\nu_\text{out}=ip$. Since $e^{-i(\omega-m\Omega_H)r_*}\approx(r-r_+)^{-ip}$ near the horizon, these roots represent the ingoing solution $\Delta^2e^{-i(\omega-m\Omega_H)r_*}$ and the outgoing solution $e^{+i(\omega-m\Omega_H)r_*}$, respectively. The ingoing solution is then given by the Frobenius series
\begin{equation}
    R_\text{in}(r,\omega)=(r-r_+)^{\nu_\text{in}}\sum_{n=0}^\infty c_n(\omega)(r-r_+)^n,\qquad c_0=1,
\end{equation}
whose coefficients obey a finite-term recurrence $P(\nu_\text{in}+n)\,c_n=F_n(c_{n-1},c_{n-2},\dots)$, with $F_n$ linear in the previous coefficients and polynomial in $\omega$. Consequently, the recurrence features simple poles at frequencies $\omega$ which satisfy
\begin{equation}
    P(\nu_\text{in}+n)=n(n+2-2ip)=-\frac{in}{\kappa_H}\Bigl(\omega-\omega_\text{MB}^{(n+2)}\Bigr)=0
\end{equation}
Since $F_n$ is linear in $c_{n-1}$, $c_{n-2}$, etc., $c_n$ carries simple poles at $\omega_\text{MB}^{(3)},\dots,\omega_\text{MB}^{(n+1)},\omega_\text{MB}^{(n+2)}$, and $R_\text{in}$ has poles at $\omega_\text{MB}^{(N)}$ for all $N\geq3$, but never at $\omega_\text{MB}^{(1)}$ or $\omega_\text{MB}^{(2)}$. Equivalently, at $\omega=\omega_\text{MB}^{(N)}$ the exponents differ by $\nu_\text{out}-\nu_\text{in}=N-2$: for $N=1$ the ingoing exponent is the larger root and for $N=2$ the roots coincide, so in both cases the ingoing series is well-defined, while for $N\geq3$ the outgoing exponent lies a positive integer $N-2$ above it, so the recursion breaks down at order $n=N-2$.

The residue of $R_\text{in}$ at $\omega_\text{MB}^{(N)}$ is accordingly proportional to the series beginning at $(r-r_+)^{\nu_\text{in}+N-2}=(r-r_+)^{\nu_\text{out}}$, i.e., to the outgoing solution, and the pole is present whenever $F_{N-2}$ does not vanish there, as is generically the case in Kerr. 
Since the source coefficients $A_j$ depend on $\omega$ only polynomially through $K$, and the spheroidal harmonic $_{-2}S_{\ell m}(a\omega;\theta)$ is analytic in $\omega$ near these frequencies, the decay of the response-filtered kernel's trailing wake at $\omega_\text{MB}^{(3)}$ arises exclusively from the poles in $R_\text{in}$ and its derivatives. The late-time Matsubara behavior of the response-filtered waveform discussed above does not originate from this pole behavior, arising instead from the near-horizon decay of the source along the trajectory, which is where the screening factors act.

\newpage
\subsection*{Kernel results for other spins}
\begin{figure}[h!]
    \centering
    \includegraphics[width=0.89\linewidth]{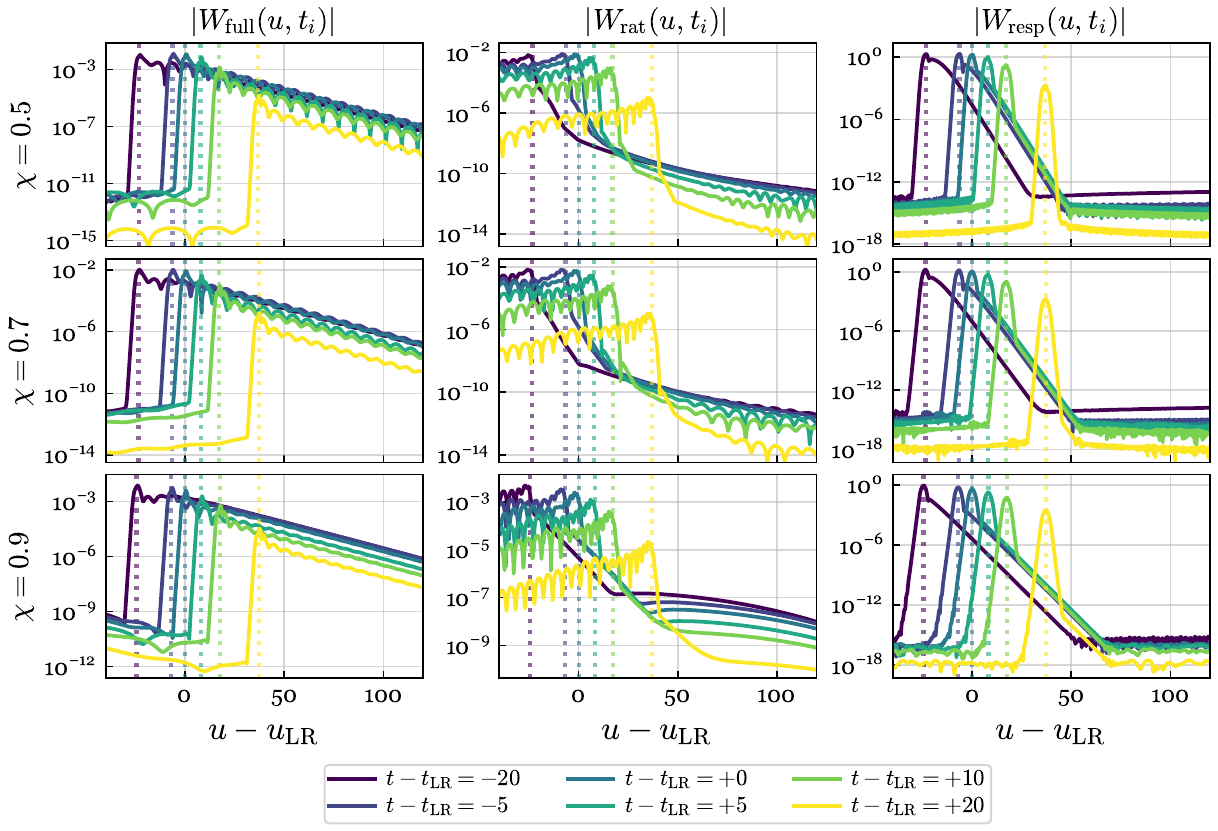}
    \caption{Slices of the unfiltered, rationally-filtered, and response-filtered kernels for fixed values of $t-t_\text{LR}$, with no normalization. The response filter localizes the contribution of each segment of the trajectory better than the rational filter for all values of $\chi$.}
    \label{fig:filtered kernel slices}
\end{figure}

\begin{figure}[h]
    \centering
    \includegraphics[width=0.49\linewidth]{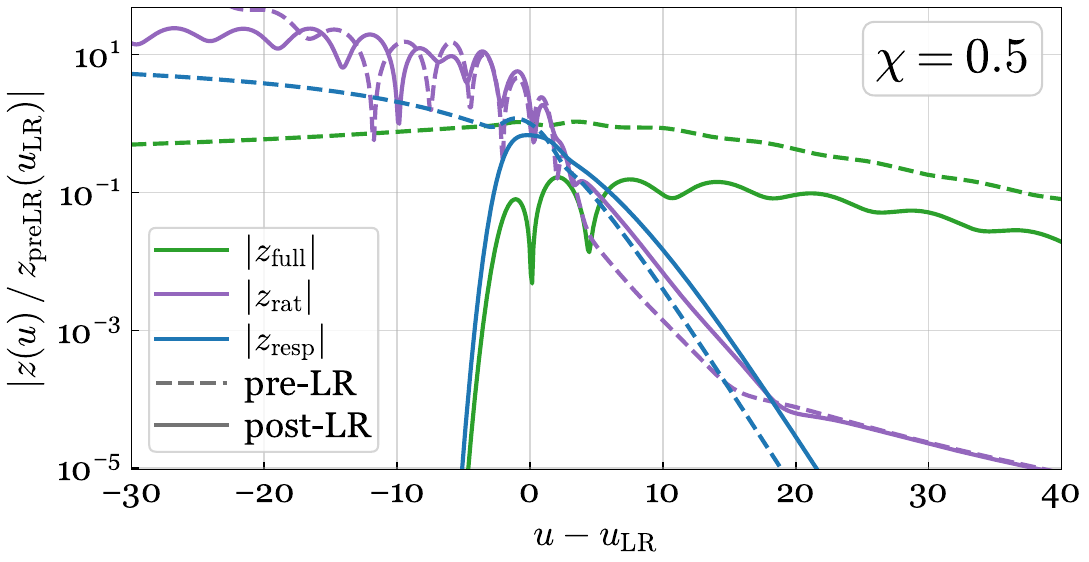}
    \hfill
    \includegraphics[width=0.49\linewidth]{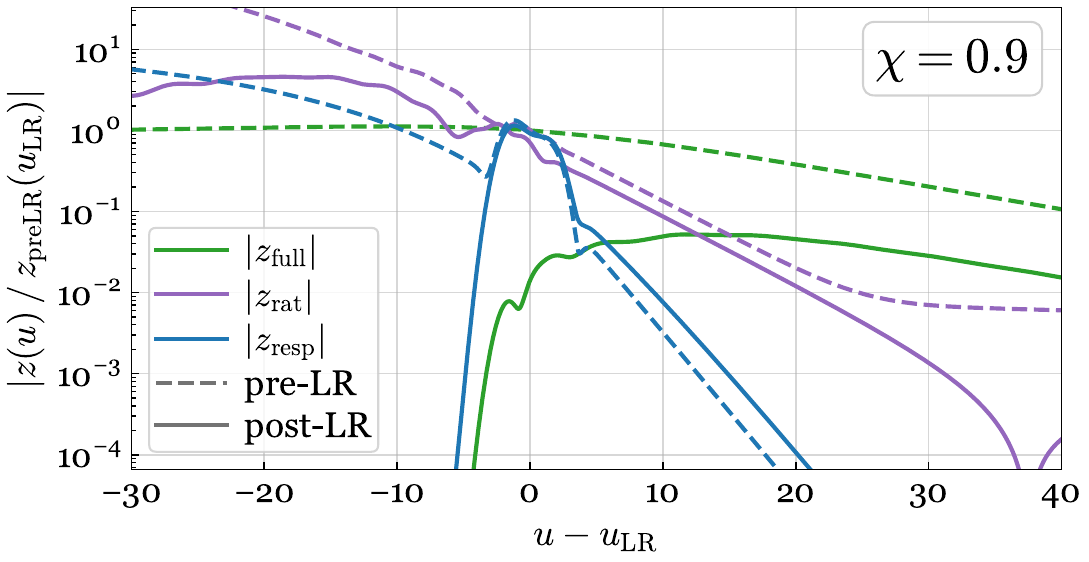}
    \caption{Identical calculations to Figure \ref{fig: lr split waveforms}, except for $\chi=0.5$ (left) and $\chi=0.9$ (right).}
    \label{fig:lr split waveforms supp}
\end{figure}

In Figure \ref{fig:filtered kernel slices}, we further develop the finding depicted in Figure \ref{fig:filtered kernels} by showing slices of the unfiltered, rationally-filtered, and response-filtered kernels at fixed values of $t-t_\text{LR}$ for $\chi\in\{0.5,0.7,0.9\}$. It is readily apparent that the enhanced ability of the response filter to localize the contribution of the source from each portion of the plunge in the waveform is not a spin-dependent statement, with this behavior demonstrated well for all three spin values. In Figure \ref{fig:lr split waveforms supp}, we reproduce Figure \ref{fig: lr split waveforms} for $\chi=0.5$ and $0.9$, showing that the response filter reveals the post-LR contribution to the waveform better than the rational filter for these spin values as well.

\subsection*{Complete complex $t$ saddle results}
\begin{figure}
    \centering
    \includegraphics[width=\linewidth]{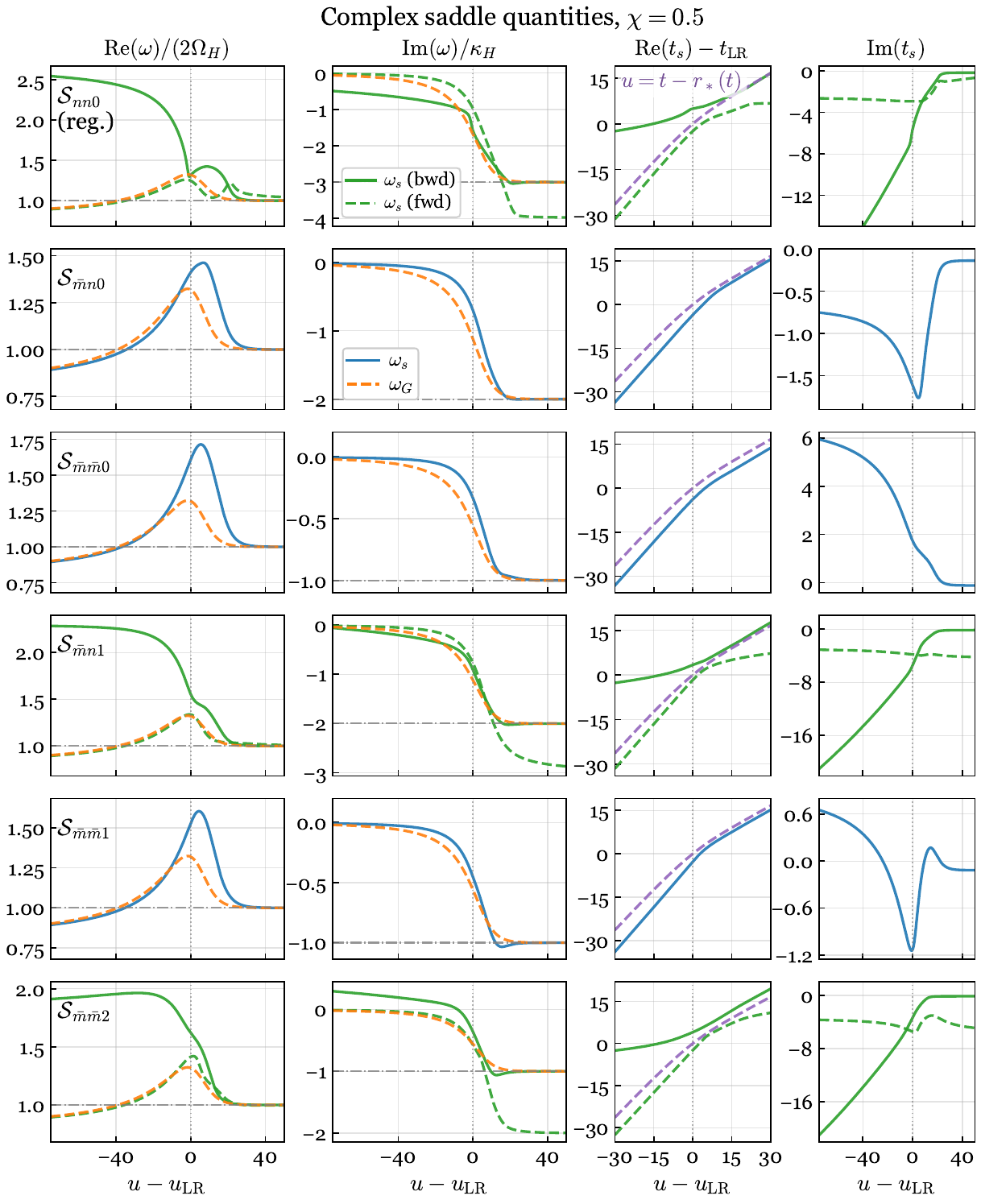}
    \caption{The complex saddle frequency $\omega_s$ and the imaginary part of the saddle time $\text{Im}(t_s)$ as a function of the resulting observer time $u$ for $\chi=0.5$, with the light-ring crossing $u_{\text{LR}}\equiv t_{\text{LR}}-r_{*,\text{LR}}$ set to zero. We also include a trace at $\omega=\omega_G$, $u=t-r_*$ for reference, and when two saddles are present, we distinguish those found by following the saddle backward (forward) from the late (early) time asymptotic saddle frequency by green solid (dashed) lines.}
    \label{fig:complex saddle quants 0.5}
\end{figure}

\begin{figure}
    \centering
    \includegraphics[width=0.95\linewidth]{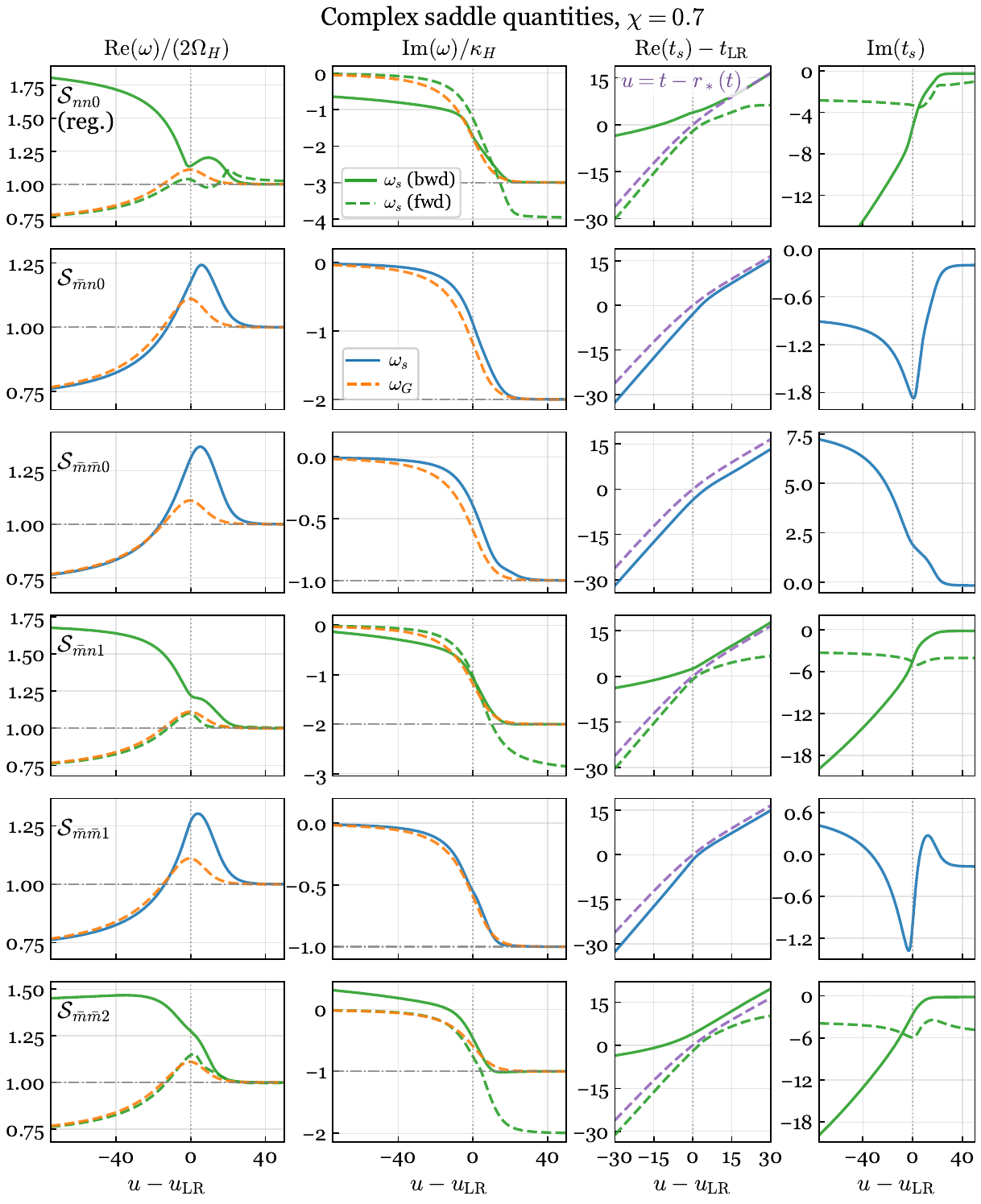}
    \caption{Identical to Figure \ref{fig:complex saddle quants 0.5}, but with $\chi=0.7$.}
    \label{fig:complex saddle quants 0.7}
\end{figure}

\begin{figure}
    \centering
    \includegraphics[width=0.95\linewidth]{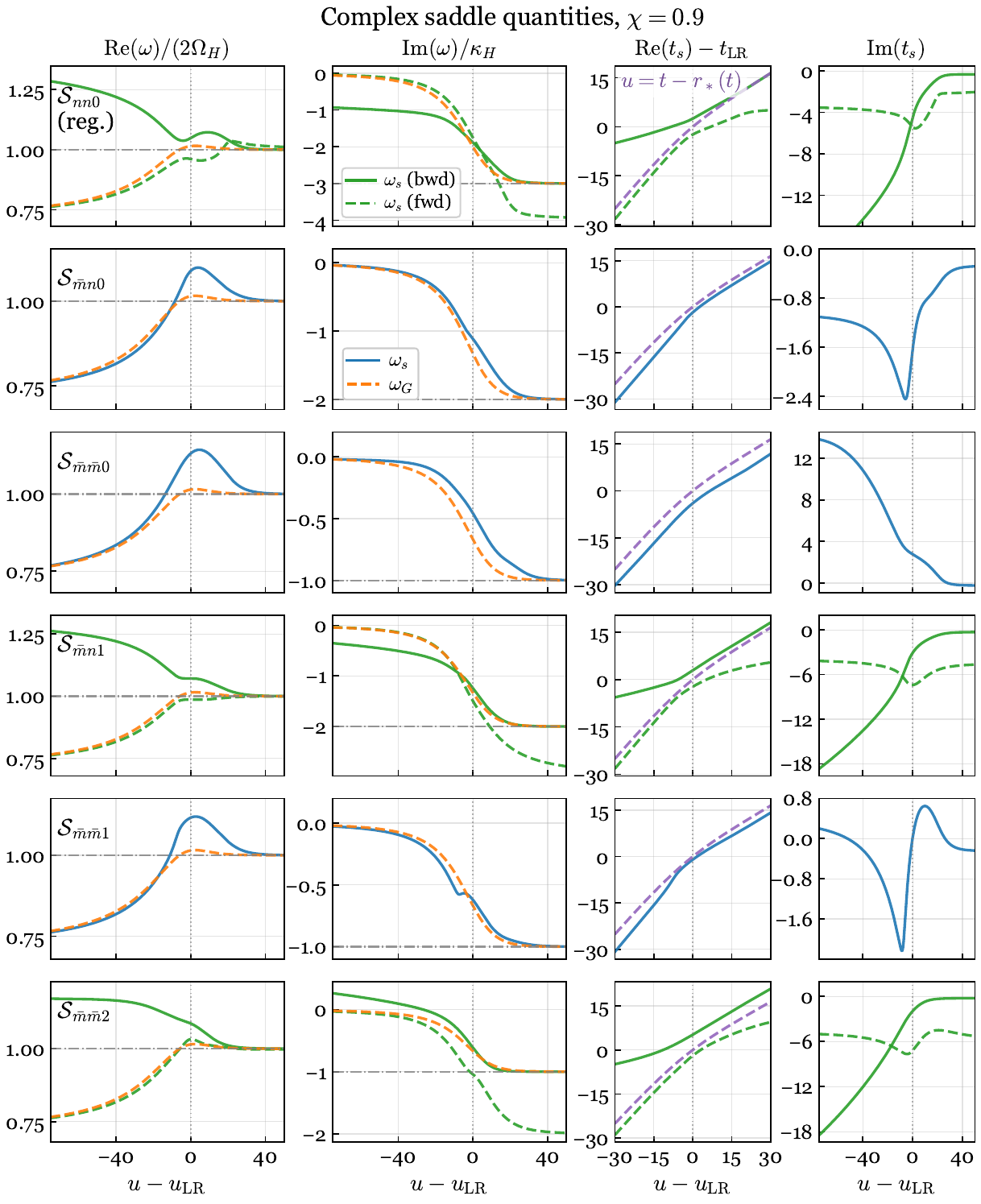}
    \caption{Identical to Figure \ref{fig:complex saddle quants 0.5}, but with $\chi=0.9$.}
    \label{fig:complex saddle quants 0.9}
\end{figure}

We present the full set of results for the three spins studied in this paper: $\chi=0.5$, $0.7$, and $0.9$. In Figs. \ref{fig:complex saddle quants 0.5}, \ref{fig:complex saddle quants 0.7}, and \ref{fig:complex saddle quants 0.9}, we present the complex frequencies and times for the saddle(s) identified for each of the six terms, noting that $S_{nn0}$ has been regularized. For some terms, a single saddle is sufficient, while others require two which have an avoided crossing near the light-ring, but in either case, both the early and late time predicted limits are reached, with a saddle matching the asymptotic trajectory of $\omega_{G,j}$. When a single saddle is found, the real part of $t_s$ roughly tracks the light-cone $u=t-r_*$, and when two saddles are found, the ``forward" saddle which matches the early-time asymptotics tracks the light-cone until the avoided crossing, when the ``backward" saddle takes over.

\begin{figure}
    \centering
    \includegraphics[width=0.95\linewidth]{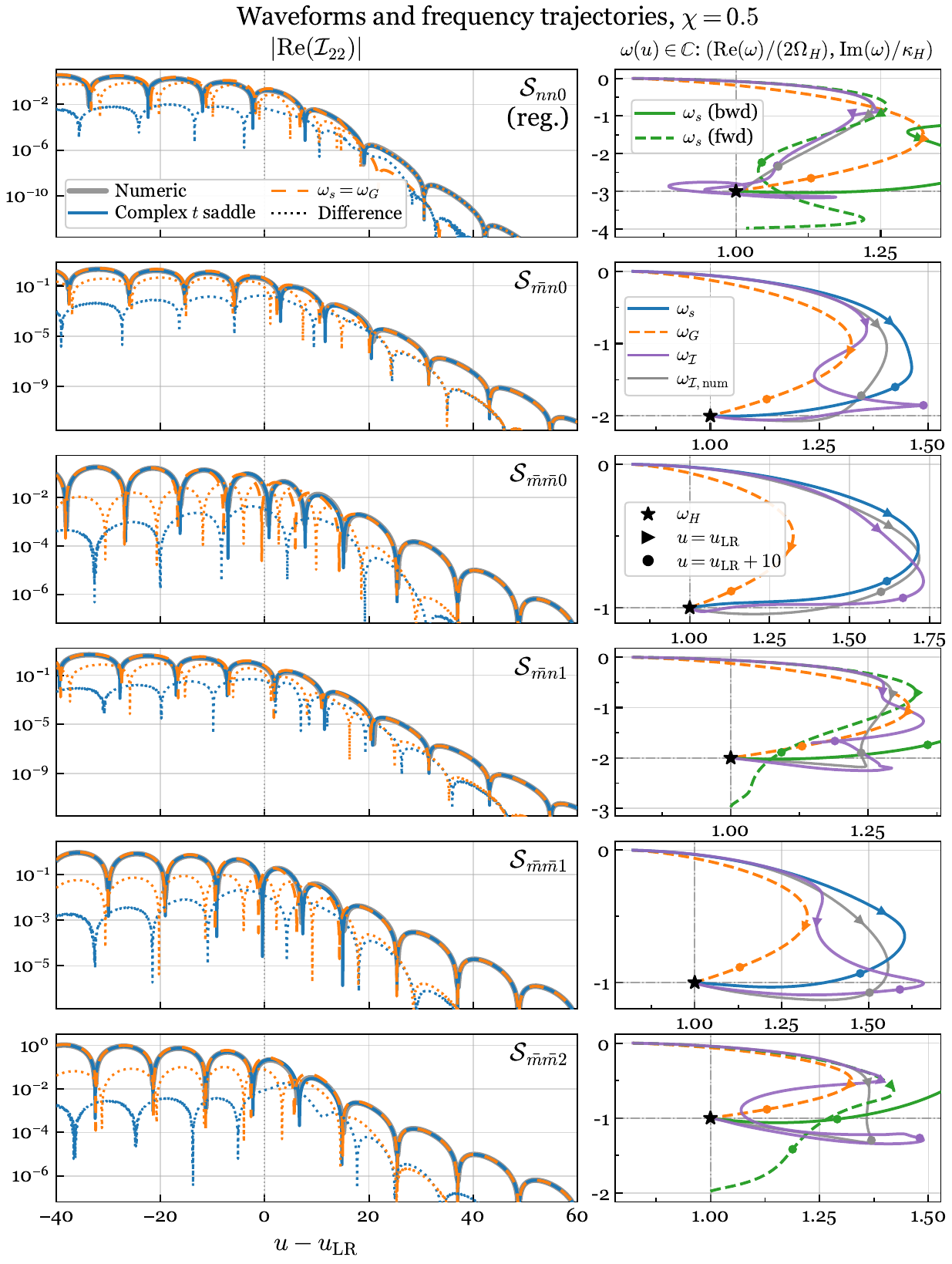}
    \caption{\textit{Left:} $\mathcal{I}_{22}(u) $ waveforms computed via the method of steepest descent. Blue and orange correspond to the analytically continued saddle and the case where $\omega_s$ is set to $\omega_G$ and $u$ to $t-r_*$, respectively. Their differences relative to a numerical waveform computed by integrating along the real frequency axis appear as dotted lines. \textit{Right:} frequency trajectories in the complex-$\omega$ plane, where $\omega_\mathcal{I}\equiv i\partial_u\mathcal{I}_{22}/\mathcal{I}_{22}$ is the instantaneous waveform frequency. Each row corresponds to one $\mathcal{S}_j$, colored markers appear at $u=u_\text{LR}$ and $u=u_\text{LR}+10$, and the star marking the expected asymptotic limit $\omega_{H,j}$.}
    \label{fig:waveform grid 0.5}
\end{figure}

\begin{figure}
    \centering
    \includegraphics[width=0.95\linewidth]{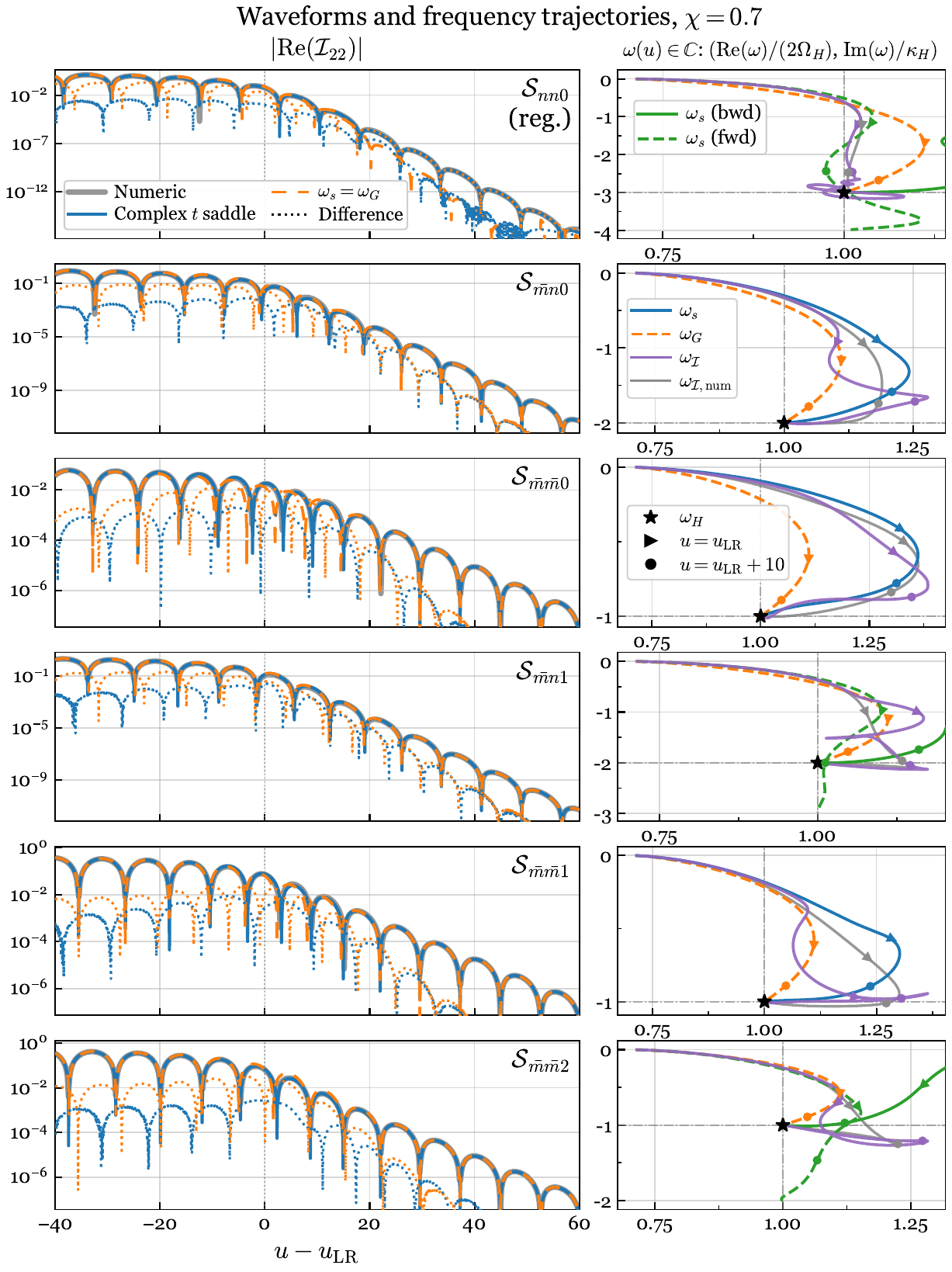}
    \caption{Identical to Figure \ref{fig:waveform grid 0.5}, but with $\chi=0.7$.}
    \label{fig:waveform grid 0.7}
\end{figure}

\begin{figure}
    \centering
    \includegraphics[width=0.95\linewidth]{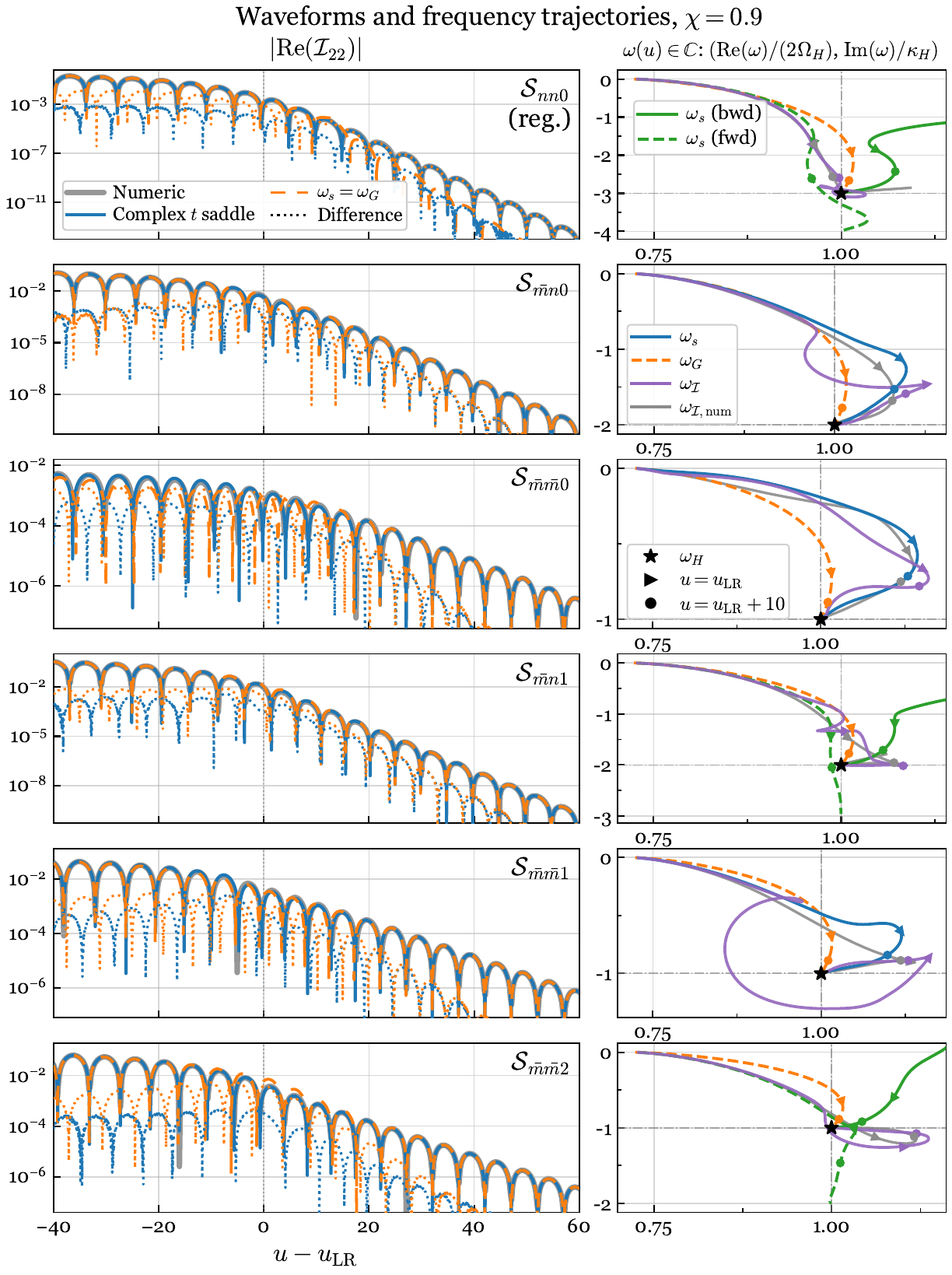}
    \caption{Identical to Figure \ref{fig:waveform grid 0.5}, but with $\chi=0.9$.}
    \label{fig:waveform grid 0.9}
\end{figure}

In Figs. \ref{fig:waveform grid 0.5}, \ref{fig:waveform grid 0.7}, and \ref{fig:waveform grid 0.9}, we plot the waveforms which are computed from the method of steepest descent using the saddle trajectories identified above. For the cases where two saddles are identified, the method of Chester, Friedman, and Ursell \cite{ChesterFriedmanUrsell} is employed to build the waveform in the vicinity of the avoided crossing -- one saddle dominates the total waveform outside of this region. We also plot the waveform computed inserting $\omega_s=\omega_G$, $u=t-r_*$ into the single-saddle steepest-descent waveform computation; this is \textit{not} equivalent to a waveform $\mathcal{I}_22(u)\sim\exp(-i\int \omega_G(u)du)$, as the evolution of the Hessian of the source phase modifies the instantaneous complex frequency of the waveform. The differences between the saddle waveforms and a ``numerical" version computed by straightforward integration of Eq. \eqref{eq: I_lm} along the real $t$ and $\omega$ axes appear as dotted curves: across all three spins considered here, the analytically continued saddle model consistently generates more accurate waveforms, though the simplified approach still performs well with the $\omega_{G,j}$ and $u=t-r_*$ inserted into the steepest-descent procedure rather than just taken as the instantaneous waveform frequency.

We also compute the evolution track of relevant frequencies in the problem: the saddle frequency $\omega_s$, the simplified model saddle frequency $\omega_G$, the instantaneous complex frequency of the steepest-descent waveform $\omega_{\mathcal{I}}\equiv i\dot{\mathcal{I}_j}/\mathcal{I}_j$, and the same frequency for the numerically-computed waveform. Various markers appear at fixed values of $u$ relative to $u_\text{LR}=t_\text{LR}-r_{*,\text{LR}}$. In general, the steepest-descent waveform complex frequency follows that of the pure numerical waveform more closely than the simplified model's $\omega_G$, even though all the curves have the same asymptotic limits (unless two saddles are identified, in which case the forward saddle matches at early times and the backward saddle at late times). The sharp deviations that $\omega_\mathcal{I}$ occasionally makes from its numerical counterpart typically emerge from imprecision in evaluating the Hessian of the source term's phase function and are generally brief in their duration.

\begin{figure}
    \centering
    \includegraphics[width=\linewidth]{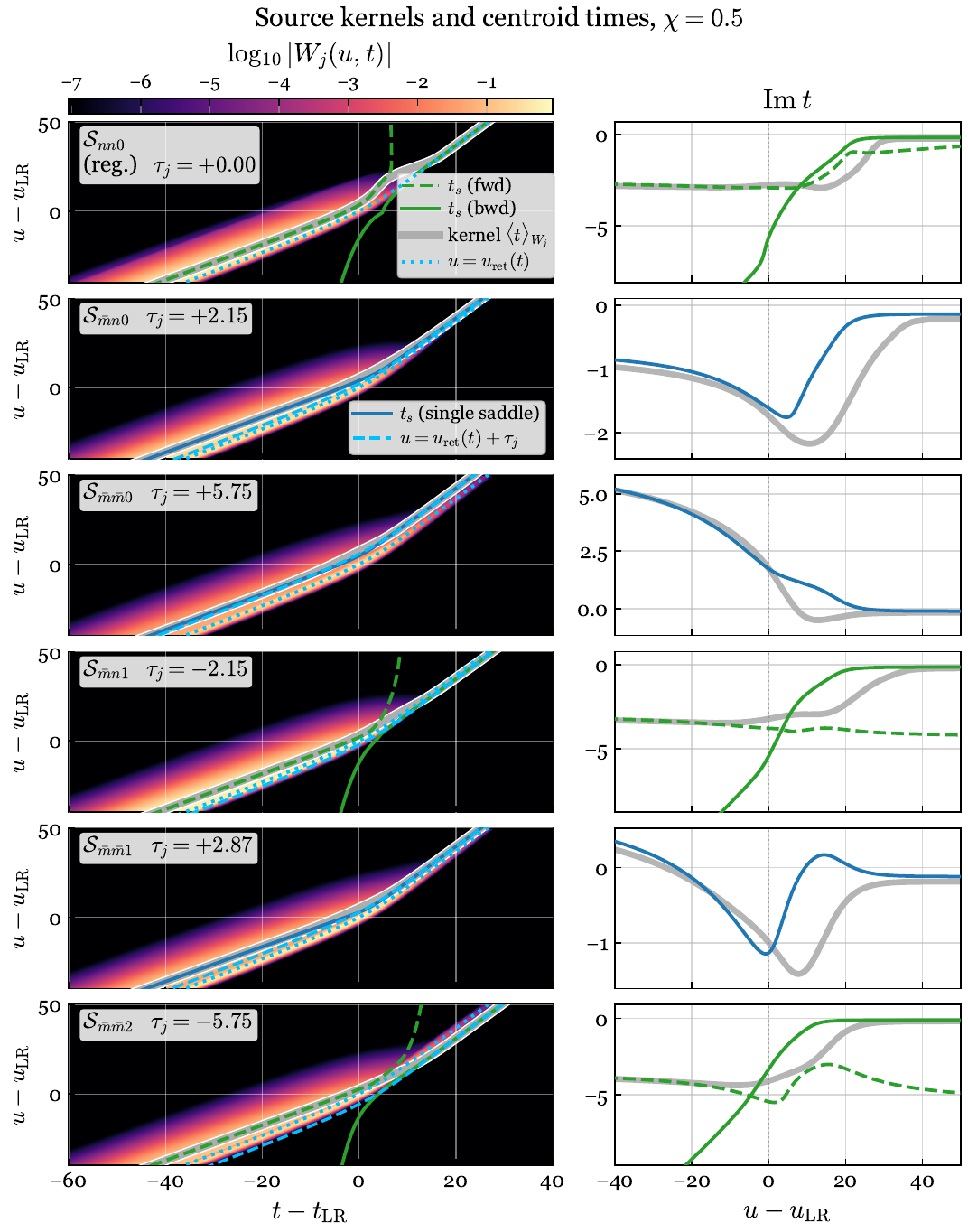}
    \caption{The magnitude of the kernel for each of the six source terms (with $S_{nn0}$ regularized to remove its pole at $\omega_\text{MB}^{(3)}$), along with the centroid and complex saddle times (real part on left, imaginary on right). Cyan lines appear on the light-cone with and without the late-time group delay $\tau_j$. The saddle times closely match the centroid before and after the light-ring crossing.}
    \label{fig:kernel quants 0.5}
\end{figure}

\begin{figure}
    \centering
    \includegraphics[width=0.95\linewidth]{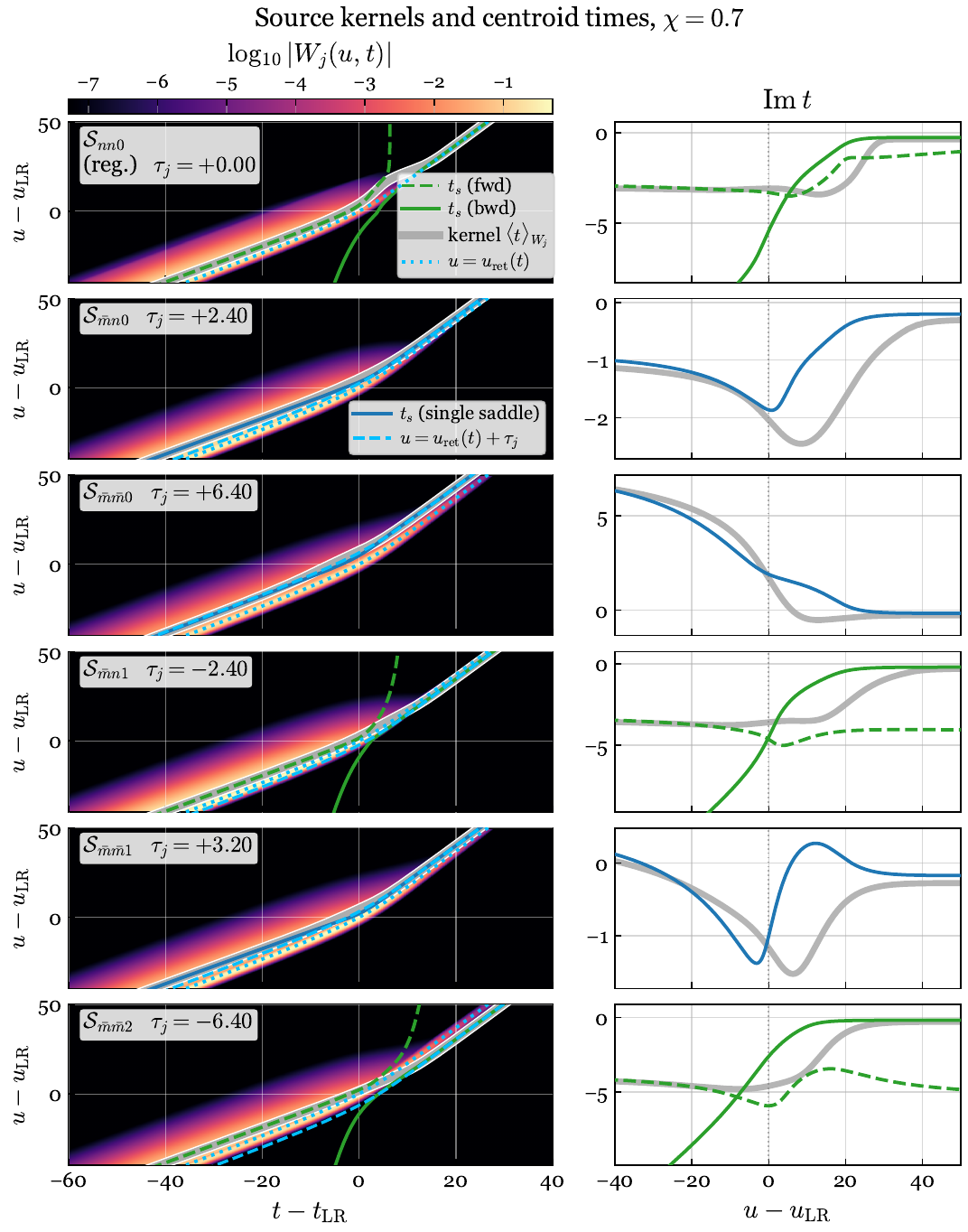}
    \caption{Identical to Figure \ref{fig:kernel quants 0.5}, but with $\chi=0.7$.}
    \label{fig:kernel quants 0.7}
\end{figure}

\begin{figure}
    \centering
    \includegraphics[width=0.95\linewidth]{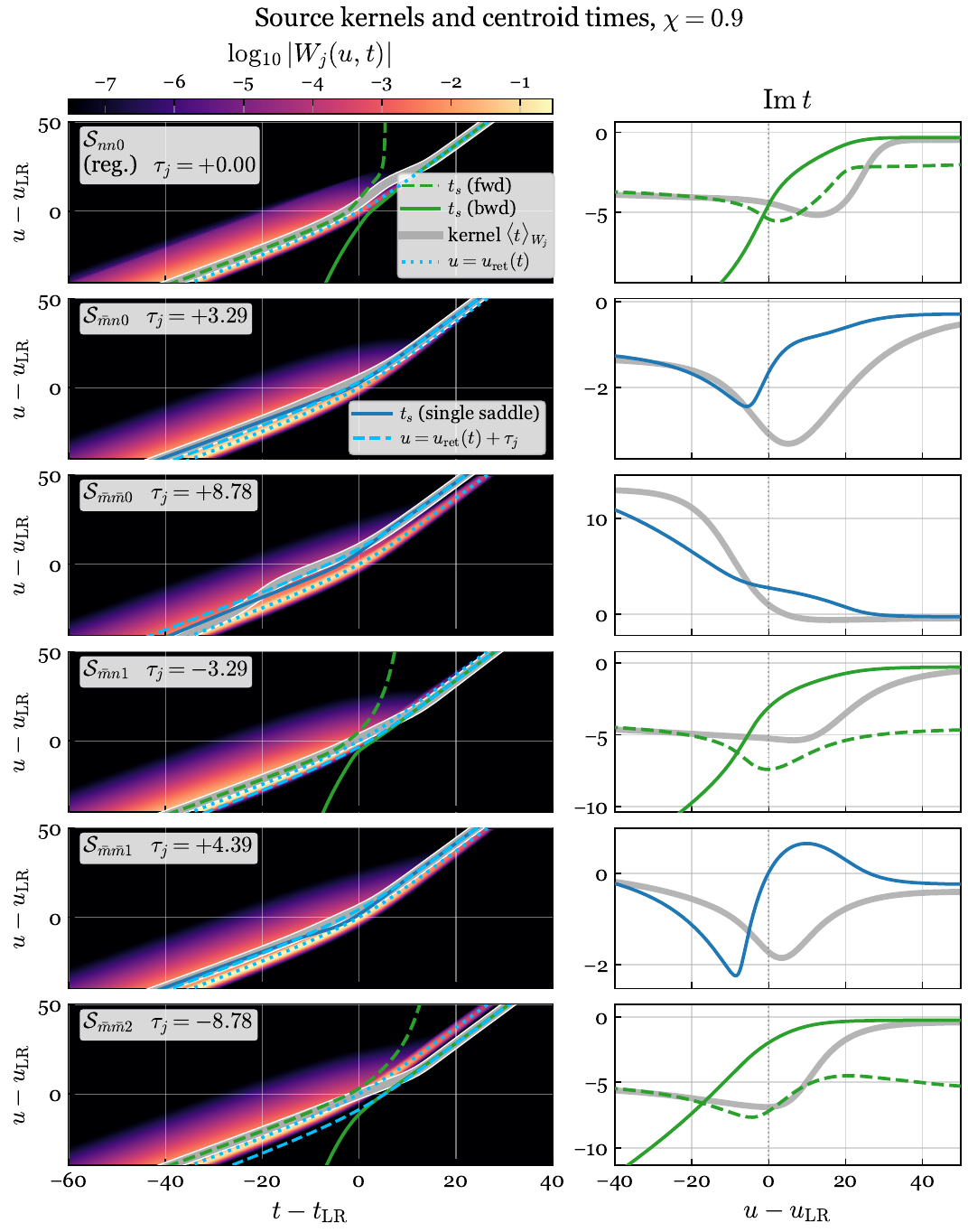}
    \caption{Identical to Figure \ref{fig:kernel quants 0.5}, but with $\chi=0.9$.}
    \label{fig:kernel quants 0.9}
\end{figure}

In Figs. \ref{fig:kernel quants 0.5}, \ref{fig:kernel quants 0.7}, and \ref{fig:kernel quants 0.9}, we compute the kernel of each of the six source terms and plot the complex saddle times $t_s$ along with the complex centroid time of the kernel. As described in the main text, the saddles match the centroid times outside of a region near the light-ring crossing, where the disagreement is small in the single saddle case or the roles of the saddles swap in the two saddle case. At late times, the structure of each source term predicts a group delay $\tau_j$, which may be positive or negative without disturbing the assertion that the kernel itself is causal on the real $t$ axis.

\begin{figure}
    \centering
    \includegraphics[width=0.99\linewidth]{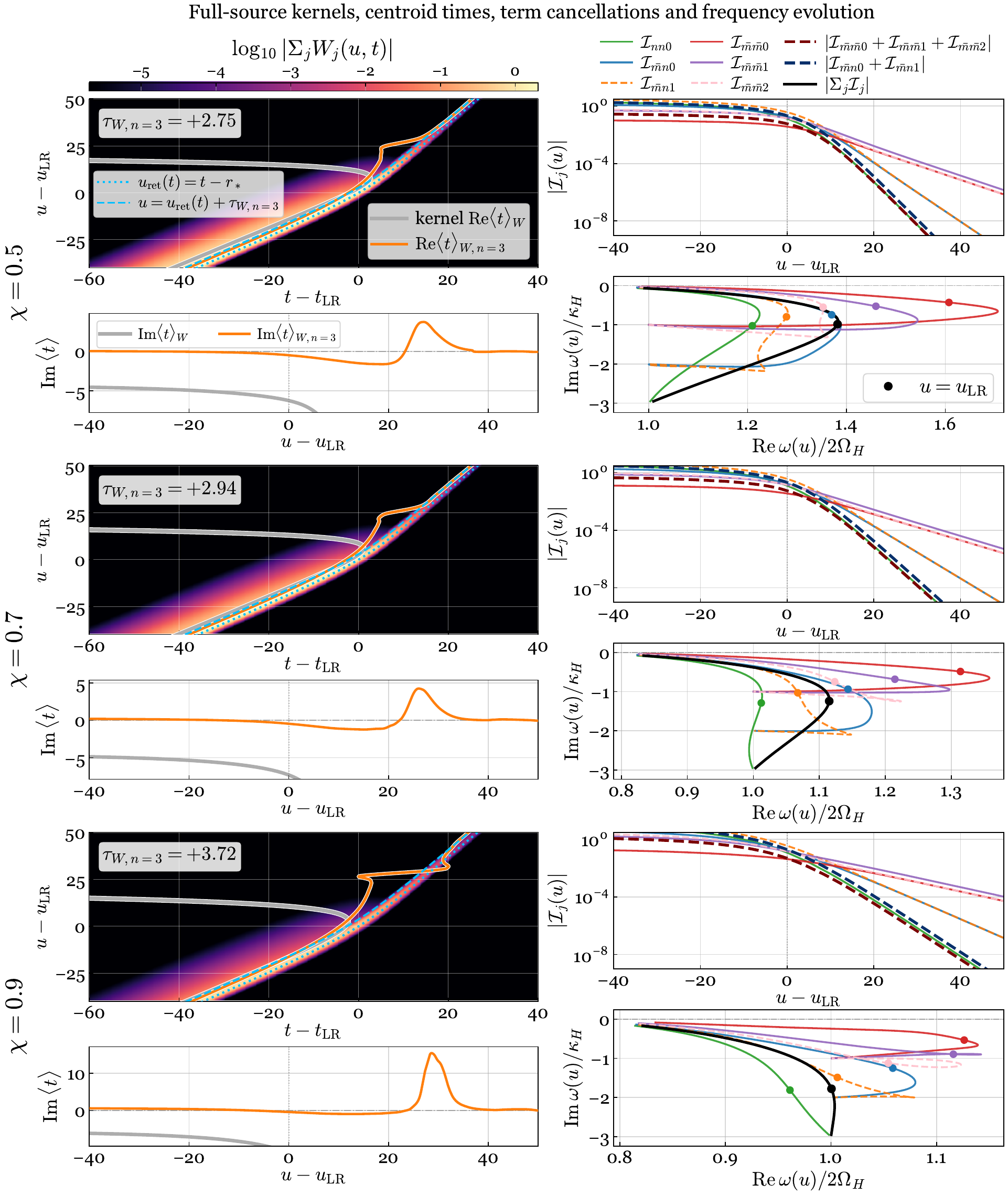}
    \caption{For each spin $\chi$, the plots are as follows. \textit{Left, top:} The kernel of the full summed source, along with the centroid of the full source (gray) and that of the isolated $\omega_\text{MB}^{(3)}$ mode expected at late times (orange). Cyan lines appear on the light-cone and with the predicted group delay $\tau_{W,n=3}$, which matches the late-time behavior of $\langle t\rangle_{W,n=3}$. \textit{Left, bottom:} The imaginary part of $\langle t\rangle_{W,n=3}$. \textit{Right, top:} The magnitude of individual source term pure-drive waveforms, along with partial sums between terms with matching late-time decay rates (thick dashed). The terms which asymptote to $\omega_\text{MB}^{(1)}$ or $\omega_\text{MB}^{(2)}$ mostly cancel after summation, leaving behind a residual decay at $\omega_\text{MB}^{(3)}$ which is captured mathematically in the near-horizon $(\omega-\omega_\text{MB}^{(1)})(\omega-\omega_\text{MB}^{(2)})$ screening factors. \textit{Right, bottom:} The instantaneous frequency of each component of the response-filtered waveform along with the full sum, demonstrating the asymptotic approach to $\omega_\text{MB}^{(3)}$.}
    \label{fig:full source kernel}
\end{figure}

Finally, in Figure \ref{fig:full source kernel}, we plot the kernel of the complete source $\sum_jW_j(u,t)$, along with the centroid time of the full source $\langle t\rangle_W$ along with that of the $\omega_\text{MB}^{(3)}$ component of the trailing wake, $\langle t\rangle_{W,n=3}$, computed as described in the main text. This centroid time tracks closely with the light-cone propagation time throughout the plunge, delayed by just a few $M$ for the majority of the trajectory. In the narrow region where the lag deviates more rapidly, the imaginary part of $\langle t\rangle_{W,n=3}$ also deviates rapidly from zero, where it remains for the rest of the plunge.

We also plot the magnitude of each source term's contribution to the total pure-drive waveform along with partial sums of the terms which have matching late-time asymptotic frequencies. These groups of terms undergo heavy self-cancellations: not only do $\mathcal{I}_{\bar mn0}$ and $\mathcal{I}_{\bar mn1}$, each of which decays at $\omega_\text{MB}^{(2)}$ at late times, sum to leave behind a residual at $\omega_\text{MB}^{(3)}$, but the same occurs for $\mathcal{I}_{\bar m\bar m0}$, $\mathcal{I}_{\bar m\bar m1}$, and $\mathcal{I}_{\bar m\bar m2}$, each of which decays at $\omega_\text{MB}^{(1)}$. These cancellations are predicted mathematically by a lowest-order expansion of the source near the horizon, which reveals the well-known $(\omega-\omega_\text{MB}^{(1)})(\omega-\omega_\text{MB}^{(2)})$ screening factors \cite{SizhengPaper1}. The evolution of the complex frequency of each term, along with that of the full sum, is also depicted, explicitly demonstrating the asymptotic approach to the expected Matsubara mode frequency.

\end{document}